\documentclass[journal]{IEEEtran}
\usepackage{lipsum,afterpage,refcount}
\usepackage{tablefootnote}
\usepackage{siunitx}
\usepackage[T1]{fontenc}
\usepackage{booktabs}
\usepackage{graphicx}
\usepackage{fancyhdr}
\usepackage{setspace}
\usepackage{url}
\usepackage{hyperref}
\usepackage{amsthm}
\usepackage{bm}
\usepackage{mathtools}
\usepackage{flexisym}
\usepackage{multirow}
\usepackage[flushleft]{threeparttable}
\usepackage{color}
\usepackage[table]{xcolor}
\usepackage{array}
\usepackage{stfloats}
\usepackage{changepage}
\usepackage{algpseudocode}
\usepackage{amssymb}
\usepackage{enumitem}
\usepackage{verbatim}
\usepackage{amsmath}
\usepackage{algorithm}
\usepackage{tikz}
\usepackage{bbding}
\usepackage{xspace}
\usepackage{academicons}
\usepackage{textcomp}
\usepackage{xparse}

\makeatletter
\global\let\tikz@ensure@dollar@catcode=\relax
\makeatother
\AtBeginDocument{
    \catcode`_=12
    \begingroup\lccode`~=`_
    \lowercase{\endgroup\let~}\sb
    \mathcode`_="8000
}

\newcolumntype{P}[1]{>{\centering\arraybackslash}p{#1}}
\newcolumntype{M}[1]{>{\centering\arraybackslash}m{#1}}
\newcolumntype{L}{>{\centering\arraybackslash}p{0.58cm}}
\newcolumntype{Z}{>{\centering\arraybackslash}p{.95cm}}
\newcolumntype{A}{>{\centering\arraybackslash}p{.8cm}}
\newcolumntype{B}{>{\centering\arraybackslash}p{.6cm}}
\newcolumntype{H}{>{\setbox0=\hbox\bgroup}c<{\egroup}@{}}

\hypersetup{
  colorlinks=true,
  linkcolor=blue!50!blue,
  citecolor=blue!50!blue,
  urlcolor=black!70!black
}

\DeclarePairedDelimiterX\set[1]\lbrace\rbrace{#1}
\DeclarePairedDelimiter{\ceil}{\lceil}{\rceil}

\newcommand*{\fun}{\NewDocumentCommand}
\AtBeginDocument{
  \catcode`_=12
  \begingroup\lccode`~=`_
  \lowercase{\endgroup\let~}\sb
  \mathcode`_="8000
}

\newcommand{\orcid}[1]{\href{https://orcid.org/#1}{\textcolor[HTML]{A6CE39}{\includegraphics[width=1.7ex]{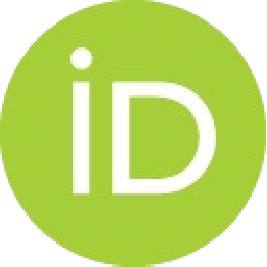}}}}

\newcommand*{\Figure}{Fig.\xspace}
\newcommand*{\figs}{Figs.\xspace}
\newcommand*{\fig}[1]{\Figure \ref{#1}}

\fun{\secn}{m}{Section \ref{#1}\xspace}
\fun{\Secn}{m}{Section \ref{#1}\xspace}

\newcommand{\eq}[1]{(\ref{#1})\xspace}
\newcommand{\eqs}{}

\newcommand*{\spc}{\text{ }}

\newcommand*{\eALMs}{118K}
\newcommand*{\eRegs}{311K}
\newcommand*{\eMems}{1782}
\newcommand*{\eDSPs}{1072}
\newcommand*{\eFreq}{388}
\newcommand*{\eResNetAGOPS}{2529}

\newcommand*{\eResNetBGOPS}{2752}

\newcommand*{\eResNetCGOPS}{2838}

\newcommand*{\x}{$\times$\xspace}
\newcommand*{\by}{$\times$}
\newcommand*{\ea}{et al.\xspace}
\newcommand*{\gx}{Arria 10 GX 1150\xspace}
\newcommand*{\sx}{Arria 10 SX 660\xspace}

\newcommand*{\lb}{\left(}
\newcommand*{\rb}{\right)}

\fun \fg{m}{\Figure \ref{#1}\xspace}
\fun{\tabThpt}{}{Throughput (GOPS) \tnote{1}}
\fun \winoConv{}{\cite{liu2021winocnn}, \cite{lavin2016fast}\xspace}
\fun \multDominant{}{\cite{liu2021winocnn}, \cite{jouppi2017datacenter}, \cite{norrie2021design}\xspace}
\fun \multComplexity{}{\cite{Lakshmi2022}, \cite{guo2019dl}, \cite{Pekmestzi1999}\xspace}

\fun \bits{mo}{\IfNoValueTF{#2}{[#1]}{}}

\fun \bld{}{\mathbf}
\fun \win{}{w}
\fun \wm{}{m}
\fun \ww{}{w}
\fun \dw{}{\win}
\fun \dm{}{\wm}
\fun \zero{}{}

\fun \nn{}{n}
\fun \one{}{}
\fun \dPsSub{}{2}

\fun \eqFpSup{O{\dw}O{\dm}o}{\bits{#1}[#3]}
\fun \eqPsSup{O{\dw}O{\dm}o}{\bits{#1}[#3]}

\fun \sffip{}{FFIP+\SMM}

\fun \MM{}{MM}

\fun \fpW{O{\win}}{#1}
\fun \psW{O{\win}}{#1}

\fun \psPrefix{}{}
\fun \fpPrefix{}{}

\fun \eqTxtAlg{ooooooooo} {\IfBooleanTF {#8}
  {\IfBooleanTF {#5}
    {\tx{{#1}}_{#2}}
    {{#1}$_{#2}$\xspace}}
  {\IfBooleanTF {#5}
    {\tx{#9{#1}}_{#2}#6[#3][#4][#7]}
    {#9{#1}$_{#2}#6[#3][#4][#7]$\xspace}}}

\fun \PsFp{oooooooo} {\IfBooleanTF {#1}
  {\eqTxtAlg[#2][#3][#4][#5][#6][\eqFpSup][#7][#8][\fpPrefix]}
  {\eqTxtAlg[#2][#3][#4][#5][#6][\eqPsSup][#7][#8][\psPrefix]}}

\fun \nx{}{\nn}
\fun \nxT{}{2}

\fun \mm{E{/?}{{\ww}{\one}}s}{\IfBooleanTF{#3}
{\alg.\MM'*/{#1}?{#2}<{}}{\alg.\MM'/{#1}?{#2}<{}}}

\fun \sm{E{/?}{{\ww}{\nx}}s}{\IfBooleanTF{#3}
{\alg.\SM'*/{#1}?{#2}<{}}{\alg.\SM'/{#1}?{#2}<{}}}

\fun \alg{E{.}{{\SMM}}t'sE{?>/<}{{\dPsSub}{\dm}}t!}
     {\PsFp[#2][#1][#4][\dW[#6][#2]][#5][#3][#7][#8]}

\fun \dW{oo} {\IfBooleanTF {#2}
    {\fpW[#1]}
    {\psW[#1]}
}

\newcommand*{\n}{\xspace}

\newcommand*{\tx}[1]{\text{#1}}

\AtBeginDocument{
  \catcode`_=12
  \begingroup\lccode`~=`_
  \lowercase{\endgroup\let~}\sb
  \mathcode`_="8000
}

\fun{\smmMacUt}{}{$\frac{\tx{\n mults} / \tx{multiplier}} {\tx{clock cycle}}$  \tnote{2}}
\fun {\smmMacUtExpl}{}{Multiplier compute efficiency, defined in \secn{sec:mu}, measures how effectively the architecture utilizes its multipliers. It can surpass 1 in \smm architectures because the observed mults/s is equal to the number of multiplications required to carry out an execution using conventional algebra divided by the measured execution time.
For prior works that did not provide this metric, the value displayed here is reverse engineered based on other provided metrics and design choices as explained further in \secn{smm:sec:results-prior-work}.}
\fun {\matSzExpl}{}{Quantifies how much smaller the minimum supported matrix sizes of a multisystolic array design are relative to a single-systolic array design with the same throughput per clock cycle roof, definition and relevance provided in \secn{smm:sec:mat-sz}.}

\fun \sAlg{E{.}{{\SMM}}t'sE{?>/<}{{}{\dm}{}{r}}t!}
     {\PsFp[#2][#1][#4][\dW[#6][#2]][#5][#3][#7][#8]}

\fun \s{}{Strassen\xspace}
\fun \sa{}{Strassen's algorithm\xspace}

\fun \citeTPU{}{\cite{jouppi2017datacenter}, \cite{norrie2021design}, \cite{jouppi2023tpu}\xspace}
\fun \citeMultiSys{}{\cite{norrie2021design}, \cite{jouppi2023tpu}, \cite{kung2019maestro}, \cite{yuzuguler2023Scale}\xspace}
\fun \citeCpu{}{\cite{sun2023accelerate}, \cite{misra2022stark}, \cite{krishnan2021multi}, \cite{khaled2020applying}, \cite{huang2020strassen}, \cite{lai2013accelerating}, \cite{lipshitz2012communication}, \cite{ballard2012communication}}

\fun \seq{}{\eqs (\ref{smm:eq:strass-first})-(\ref{smm:eq:strass-last})\xspace}
\fun \seqA{}{\eq{smm:eq:strass-first}\xspace}
\fun \seqB{}{\eq{smm:eq:strass-last}\xspace}

\fun \An{sE{?'}{{ij}{}}}{\m.A?{#2}'{#3}}
\fun \Bn{sE{?'}{{ji}{}}}{\m.B?{#2}'{#3}}
\fun \Tn{sE{?'}{{x} {}}}{\m.T?{#2}'{#3}}
\fun \Sn{sE{?'}{{x} {}}}{\m.S?{#2}'{#3}}
\fun \Qn{sE{?'}{{x} {}}}{\m.Q?{#2}'{#3}}
\fun \Cn{sE{?'}{{ij}{}}}{\m.C?{#2}'{#3}}

\fun \Anv{sE{?'}{{ij}{i:m:,:}}}{\m.A?{#2}'{#3}}
\fun \Bnv{sE{?'}{{ji}{:,j:n:}}}{\m.B?{#2}'{#3}}
\fun \Tnv{sE{?'}{{x} {i:m:,:}}}{\m.T?{#2}'{#3}}
\fun \Snv{sE{?'}{{x} {:,j:n:}}}{\m.S?{#2}'{#3}}
\fun \Qnv{sE{?'}{{x} {i:k:,:}}}{\m.Q?{#2}'{#3}}
\fun \Cnv{sE{?'}{{ij}{i:k:,:}}}{\m.C?{#2}'{#3}}

\fun \Av{sE{?'}{{}{i:m:,:}}}{\m.A?{#2}'{#3}}
\fun \Bv{sE{?'}{{}{:,j:n:}}}{\m.B?{#2}'{#3}}
\fun \Tv{sE{?'}{{}{i:m:,:}}}{\m.T?{#2}'{#3}}
\fun \Sv{sE{?'}{{}{:,j:n:}}}{\m.S?{#2}'{#3}}
\fun \Qv{sE{?'}{{}{i:m:,:}}}{\m.Q?{#2}'{#3}}
\fun \Cv{sE{?'}{{}{i:m:,:}}}{\m.C?{#2}'{#3}}

\fun \m{sE{.?'}{{}{}{}}}{\IfBooleanTF {#1} {\bld{#2_{#3}}_{#4}} {$\bld{#2_{#3}}_{#4}$\xspace}}
\fun \A{s}{\IfBooleanTF {#1} {\bld{A}} {$\bld{A}$\xspace}}
\fun \B{s}{\IfBooleanTF {#1} {\bld{B}} {$\bld{B}$\xspace}}
\fun \AB{s}{\IfBooleanTF {#1} {\bld{AB}} {$\bld{A}$ and $\bld{B}$\xspace}}
\fun \C{s}{\IfBooleanTF {#1} {\bld{C}} {$\bld{C}$\xspace}}
\fun \T{s}{\IfBooleanTF {#1} {\bld{T}} {$\bld{T}$\xspace}}
\fun \Ss{s}{\IfBooleanTF {#1} {\bld{S}} {$\bld{S}$\xspace}}

\fun \SMM{}{SMM}
\fun \smm{E{/?}{{\ww}{r}}s}{\IfBooleanTF{#3}
  {\alg.\SMM'*/{#1}?{#2}<{}}{\alg.\SMM'/{#1}?{#2}<{}}}
\fun \smmArch{E{/?}{{\win}{r}}}{\alg.\SMM/{#1}?{#2}<{}}
\fun \mmArch{E{/?}{{\win}{r}}}{\alg.\MM/{#1}?{#2}<{}}
\fun{\macUtA}{}{$\tx{roof}\lb\frac{\tx{\n mults} / \tx{multiplier}} {\tx{clock cycle}}\rb$ \xspace}

\fun \block{}{block\xspace}
\fun \blocks{}{blocks\xspace}
\fun \subblock{}{sub-block\xspace}
\fun \subblocK{}{sub-block}
\fun \subblocks{}{sub-blocks\xspace}
\fun \gemmtile{}{GEMM tile\xspace}
\fun \gemmtiles{}{GEMM tiles\xspace}

\fun \minSzMetric{}{$\tx{roof}\lb\frac{\tx{mults}/\tx{clock cycle}}{\tx{min. mat. size (h\by w)}}\rb$ \xspace}

\newcommand*{\footNoteRefs}{\item[1-2] See the corresponding definitions from Table \ref{smm:tab:second}.}

\newcommand\copyrighttext{%
  \scriptsize \textcopyright
  2025 IEEE. Personal use of this material is permitted. Permission
  from IEEE must be obtained for all other uses, in any current or future
  media, including reprinting/republishing this material for advertising or
  promotional purposes, creating new collective works, for resale or
  redistribution to servers or lists, or reuse of any copyrighted
  component of this work in other works.
  Accepted for publication in IEEE Transactions on Very Large Scale Integration (VLSI) Systems. DOI: 10.1109/TVLSI.2025.3530785}
\newcommand\copyrightnotice{%
  \begin{tikzpicture}[remember picture,overlay]
    \node[anchor=south,yshift=0pt] at (current page.south) {\fbox{\parbox{\dimexpr\textwidth-\fboxsep-\fboxrule\relax}{\copyrighttext}}};
  \end{tikzpicture}}

\begin{document}

\title{Strassen Multisystolic Array Hardware Architectures}

\author{Trevor~E.~Pogue~\orcid{0000-0002-6791-3758} and Nicola~Nicolici~\orcid{0000-0001-6345-5908},~\IEEEmembership{Senior Member,~IEEE}
  \IEEEcompsocitemizethanks{\IEEEcompsocthanksitem T. E. Pogue and N. Nicolici are with the Department of Electrical and Computer Engineering, McMaster University, Hamilton, ON, L8S 4L8, Canada \protect\\
    Email: poguete@mcmaster.ca; nicolici@mcmaster.ca}}

\IEEEpeerreviewmaketitle

\IEEEtitleabstractindextext{%
\begin{abstract}
    While Strassen's matrix multiplication algorithm reduces the complexity of naive matrix multiplication, general-purpose hardware is not suitable for achieving the algorithm's promised theoretical speedups.
This leaves the question of if it could be better exploited in custom hardware architectures designed specifically for executing the algorithm.
However, there is limited prior work on this and it is not immediately clear how to derive such architectures or if they can ultimately lead to real improvements.
We bridge this gap, presenting and evaluating new systolic array architectures that efficiently translate the theoretical complexity reductions of \sa directly into hardware resource savings.
Furthermore, the architectures are multisystolic array designs that can multiply smaller matrices with higher utilization than single-systolic array designs.
The proposed designs implemented on FPGA reduce DSP requirements by a factor of $1.14^r$ for $r$ implemented Strassen recursion levels, and otherwise require overall similar soft logic resources when instantiated to support matrix sizes down to 32\by32 and 24\by24 at 1-2 levels of Strassen recursion, respectively.
We evaluate the proposed designs both in isolation and in an end-to-end machine learning accelerator compared to baseline designs and prior works, achieving state-of-the-art performance.

\end{abstract}
}

\maketitle
\IEEEdisplaynontitleabstractindextext
\IEEEpeerreviewmaketitle
\section{Introduction}
\copyrightnotice
\IEEEPARstart{D}{ue} to the rising demand for optimized hardware acceleration of general matrix multiplication (GEMM), the field of hardware design continues to see innovation for ways of better exploiting the inherent parallelism to speed up computation.
However, at a certain point, after technology scaling slows to a halt and the system-level optimizations and known parallelism are exhausted, an accelerator wall exists which limits further progress on the implementation side \cite{fuchs2019accelator}.
A less-explored path for advancement beyond this wall is through reducing the computation at the algebraic level, by computing the same output from a re-arranged compute pattern requiring fewer or cheaper operations to be executed in hardware.

One of the area-dominant computational resources in GEMM and deep learning accelerators can commonly be the multiply-accumulate (MAC) units \multDominant, and an accelerator's throughput can be directly limited by how many MAC units can be afforded in its hardware budget.
As a result, surpassing this performance per MAC limit has been focused on recently with minimal filtering algorithms applied to convolutional neural networks \winoConv and with application of fast inner-product algorithms for speeding up deep learning and GEMM workloads \cite{pogue2024fast}.

The Strassen matrix multiplication algorithm \cite{strassen1969gaussian} can also theoretically be used to reduce the complexity of naive matrix multiplication.
However, its execution on general-purpose central processing units (CPU)s and graphics processing units (GPU)s has been shown to be not suitable for achieving the algorithm's promised theoretical speedups \citeCpu.
\sa even increases execution time on CPUs/GPUs unless the matrix widths/heights are in the range of at least 1024 elements or larger.
This limits the benefits of using the algorithm on these devices for modern workloads that do not decompose to such large matrix multiplications.
\sa contains hidden overheads such as extra data accesses required for reading/computing/storing additional intermediate matrices before/after the matrix multiplication steps.
These extra steps all add to the overall execution time beyond what is expected from a theoretical analysis based on the number of arithmetic operations performed alone.

This then leaves questions surrounding if the promised theoretical complexity reductions can be more efficiently achieved in custom hardware architectures designed specifically for executing \sa.
However, prior work on this topic is limited and it is not immediately clear how to design such architectures or if they can truly lead to real improvements.
In this work, we bridge this gap by presenting and evaluating new systolic array hardware architectures for efficiently exploiting \sa.
The proposed architectures achieve a more efficient implementation of \sa compared to what is possible through execution on CPUs and GPUs by pipelining and performing the extra data movement and addition steps at all levels of recursion in parallel with the matrix multiplications.
The Strassen architectures are functionally equivalent to conventional multisystolic array designs while allowing the theoretical complexity reductions of \sa to be translated directly into hardware resource savings, even for multiplication of small matrices.
Furthermore, the architectures are multisystolic array designs, which is a type of design that can multiply smaller matrices with higher utilization than a single-systolic array design.

Compared to a conventional multisystolic array design, the proposed architecture implemented on FPGA uses 1.3\x fewer DSP units and a similar amount of soft logic resources when instantiated for multiplying matrix sizes down to 24\by24 at 2 levels of Strassen recursion.
We demonstrate how the proposed systolic array architectures are able to increase conventional multiplications/multiplier/clock cycle limits while also allowing the design to scale up in size without increasing the minimum supported matrix sizes.

\section{Background and Related Work}
\label{smm:sec:background}
\subsection{Conventional Matrix Multiplication}
A conventional matrix multiplication algorithm
computes $\bld{C} = \bld{A} \bld{B}$ for $\bld{A}$ of size $M$\by$K$ and $\bld{B}$ of size $K$\by$N$, where each element $c_{i,j}$ of $\bld{C}$ is calculated as follows:
\begin{align} \label{smm:eq:mmZero}
  c_{i,j} = \sum_{k=1}^{K} a_{i,k} b_{k,j} \,.
\end{align}

Alternatively, \C can also be computed by dividing \A and \B into 4 matrix \blocks, where \C is then computed by carrying out 8 matrix \block multiplications and 4 matrix \block additions between the \A and \B \blocks as follows:
\begin{align}
  \label{smm:eq:mm}
  \begin{bmatrix}
    \bld{C_{11}} & \bld{C_{12}} \\
    \bld{C_{21}} & \bld{C_{22}}
  \end{bmatrix}{=}\begin{bmatrix}
    \bld{A_{11}B_{11}}{+}\bld{A_{12}B_{21}} & \bld{A_{11}B_{12}}{+}\bld{A_{12}B_{22}}\\
    \bld{A_{21}B_{11}}{+}\bld{A_{22}B_{21}} & \bld{A_{21}B_{12}}{+}\bld{A_{22}B_{22}} \end{bmatrix} .
\end{align}
This process can then be carried out recursively again for each matrix \block product by splitting the matrix \blocks again into smaller \blocks and repeating the same process.

\subsection{Strassen Matrix Multiplication}
Strassen's fast matrix multiplication algorithm \cite{strassen1969gaussian}
provides a way to carry out \eq{smm:eq:mm} instead using 7 matrix \block multiplications and 18 matrix \block additions as follows:
\begin{align}
  \label{smm:eq:strass-first}
  \begin{split}
    \bld{T_{1}} &= \bld{A_{11}} + \bld{A_{22}}  \\
    \bld{T_{2}} &= \bld{A_{21}} + \bld{A_{22}}  \\
    \bld{T_{3}} &= \bld{A_{11}}  \\
    \bld{T_{4}} &= \bld{A_{22}}  \\
    \bld{T_{5}} &= \bld{A_{11}} + \bld{A_{12}}  \\
    \bld{T_{6}} &= \bld{A_{21}} - \bld{A_{11}}  \\
    \bld{T_{7}} &= \bld{A_{12}} - \bld{A_{22}}
  \end{split}
  \qquad\qquad
  \begin{split}
    \bld{S_{1}} &= \bld{B_{11}} + \bld{B_{22}}  \\
    \bld{S_{2}} &= \bld{B_{11}}  \\
    \bld{S_{3}} &= \bld{B_{12}} - \bld{B_{22}}  \\
    \bld{S_{4}} &= \bld{B_{21}} - \bld{B_{11}}  \\
    \bld{S_{5}} &= \bld{B_{22}}  \\
    \bld{S_{6}} &= \bld{B_{11}} + \bld{B_{12}}  \\
    \bld{S_{7}} &= \bld{B_{21}} + \bld{B_{22}}
  \end{split}
\end{align}
\begin{align}
  \begin{split}
    \bld{Q_{1}} &= \bld{T_{1}} \cdot \bld{S_{1}} \\
    \bld{Q_{2}} &= \bld{T_{2}} \cdot \bld{S_{2}} \\
    \bld{Q_{3}} &= \bld{T_{3}} \cdot \bld{S_{3}} \\
    \bld{Q_{4}} &= \bld{T_{4}} \cdot \bld{S_{4}} \\
    \bld{Q_{5}} &= \bld{T_{5}} \cdot \bld{S_{5}} \\
    \bld{Q_{6}} &= \bld{T_{6}} \cdot \bld{S_{6}} \\
    \bld{Q_{7}} &= \bld{T_{7}} \cdot \bld{S_{7}} \\
  \end{split}
  \qquad\qquad
  \begin{split}
    \label{smm:eq:strass-last}
    \bld{C_{11}} &= \bld{Q_1} + \bld{Q_4} - \bld{Q_5} + \bld{Q_7} \\
    \bld{C_{12}} &= \bld{Q_3} + \bld{Q_5} \\
    \bld{C_{21}} &= \bld{Q_2} + \bld{Q_4} \\
    \bld{C_{22}} &= \bld{Q_1} - \bld{Q_2} + \bld{Q_3} + \bld{Q_6} \\
  \end{split} \,.
\end{align}
Similarly to \eq{smm:eq:mm}, this algorithm can also be repeated recursively for each matrix \block multiplication, leading to an asymptotic complexity reduction compared to conventional matrix multiplication algorithms such as \eq{smm:eq:mmZero} and \eq{smm:eq:mm}.

\subsubsection{Winograd Form}
The Winograd form of the Strassen algorithm \cite{winograd1971multiplication} has the same asymptotic complexity but requires 15 matrix \block additions at each level of recursion rather than 18.
However, for fixed-point data types, this form increases the multiplier input datapath bitwidth by up to 2 bits for each recursion level implemented rather than 1 bit, which reduces the implementation benefits.
Due to this, we focus on the original form of the Strassen algorithm from \seq in our work instead.

\subsection{Prior Work on Multisystolic Array Systems}
\label{multi-sys}
Systolic arrays, which we also refer to as matrix multiplication units (MXU)s for convenience, are an effective choice for use in GEMM accelerators as they significantly reduce the required memory traffic and can reach high clock frequencies due to their short and regular interconnects.
Systolic array architectures have been used in state-of-the-art GEMM and deep learning accelerators such as the Tensor Processing Unit (TPU) \citeTPU, among others \cite{pogue2024fast}, \cite{zhang2019caffeine}.
However, a systolic array can only be fully utilized when the input matrix sizes at minimum match the dimensions of the systolic array or are larger, and real workloads have limits to the matrix sizes being multiplied.

There is then a limit to how fast the workload can be accelerated on a single-systolic array design.
This is because, even if more compute resources are instantiated to scale up the size of the systolic array, the systolic array will begin to be underutilized after its size surpasses the workload's matrix sizes, and the workload will not be able to execute any faster.
This is particularly true in modern workloads such as deep learning acceleration, where the matrix sizes that the workloads break down to can be smaller than the maximum systolic array size that could be instantiated in an accelerator \citeMultiSys.

To combat this, multiple smaller systolic arrays can be used in parallel, which allows for the total compute power in the systolic array system to increase while the minimum supported matrix sizes remain the same.
Prior works \cite{kung2019maestro}, \cite{yuzuguler2023Scale} achieve this by implementing variations of \eq{smm:eq:mm} by dividing larger matrices into smaller matrix \blocks, executing the smaller matrix \block multiplications on multiple smaller systolic arrays.
The \block products are then later summed up to form the final larger matrix multiplication product.
In this work, we show how to efficiently implement \seq in hardware to achieve this same goal with less hardware resources.

\subsection{Prior Work on Executing Strassen on CPUs and GPUs}
\sa has been well explored in prior work for execution on general-purpose CPUs and GPUs \citeCpu.
However, its execution on CPUs and GPUs in these prior works is unable to efficiently achieve the algorithm's promised theoretical speedups unless the widths/heights of the matrices being multiplied are in the range of at least 1024 elements or even much larger.

This non-optimal execution of \sa in CPUs and GPUs stems from irregularities introduced in the algorithm such as extra data accesses required for reading/computing/storing additional intermediate matrices before/after the matrix multiplication steps.
These irregularities all add to the overall execution time beyond what would be expected purely from a theoretical analysis of only the number of required arithmetic operations \cite{krishnan2021multi}, \cite{lipshitz2012communication}.

\subsubsection{Theoretical Complexity Reductions of Strassen's Algorithm}
\label{smm:sec:theoretical-complexity}
In this subsection, we establish what the expected theoretical complexity reductions of Strassen's algorithm are based on number of operations, and how the achieved speedups in prior works on CPU/GPU Strassen implementations fall short of achieving these theoretical complexity reductions.

Letting $M = N = K = n$, the complexity of \sa in number of arithmetic operations is $\mathcal{O}\lb n^{2.8074}\rb$ \cite{strassen1969gaussian}.
Conventional matrix multiplication \eq{smm:eq:mmZero} requires $n^3$ multiplications and $n^2\lb n-1 \rb$ additions for the following number of total operations:
\begin{align}
  \label{smm:eq:mm-ops}
  n^3 + n^2\lb n-1 \rb \,.
\end{align}
In contrast, \sa \eq{smm:eq:strass-first} for 1 recursion level requires $7n^3/8$ multiplications and $7\spc n^2\lb n/2-1 \rb/4 + 18\spc n^3/8$ additions for the following number of total operations:
\begin{align}
  \label{smm:eq:smm-ops}
  7n^3/8 + 7\spc n^2\lb n/2-1 \rb/4 + 18\spc n^3/8 \,.
\end{align}
The Winograd form of Strassen's algorithm \cite{winograd1971multiplication} for 1 recursion level requires $7n^3/8$ multiplications and $7\spc n^2\lb n/2-1 \rb/4 + 15\spc n^3/8$ additions for the following number of total operations:
\begin{align}
  \label{smm:eq:swmm-ops}
  7n^3/8 + 7\spc n^2\lb n/2-1 \rb/4 + 15\spc n^3/8 \,.
\end{align}
By comparing \eq{smm:eq:mm-ops} to \eq{smm:eq:smm-ops} and \eq{smm:eq:swmm-ops} for different values of $n$ we can then see that \sa requires fewer operations than conventional matrix multiplication for matrix sizes of $n \ge 16$, and $n \ge 13$ for the Winograd form of \sa.

However, \sa on CPUs and GPUs in prior works only starts providing some speedups over traditional matrix multiplication for matrix sizes $n$ of at least 20000 \cite{sun2023accelerate}, 16384 \cite{misra2022stark}, 896 \cite{krishnan2021multi}, 5000 \cite{khaled2020applying}, 1536 \cite{huang2020strassen}, 1006 \cite{lai2013accelerating}, and 1000 \cite{lipshitz2012communication} \cite{ballard2012communication}.
This limits the applicability of \sa on CPUs and GPUs for modern workloads such as deep learning that do not always decompose to such large matrix multiplications.

As derived above, prior works on CPU/GPU implementations require matrix sizes of at least 896-16384 before having benefits rather than the much lower theoretical threshold of 13 or 16.
In contrast, the custom Strassen hardware architectures presented in this work translate the benefits of \sa into hardware resource savings rather than reductions in execution time.
The proposed designs more closely achieve the theoretical complexity reductions of \sa compared to prior works on CPU/GPU implementations.
This is demonstrated in our results through the fact that the proposed architectures present area savings while achieving the same throughput/clock cycle as traditional designs even when instantiated for multiplying matrices down to size 24\by24.
Additionally, for $r$ Strassen recursion levels implemented, the proposed designs achieve $(8/7)^r$ times reduction in multipliers as expected from \seq compared to conventional designs without significant increase in other hardware components or any increase in throughput/clock cycle.

\subsection{Prior Work on Custom Strassen Hardware Architectures}
\label{background:sec:smm}
While software implementations of \sa on CPUs and GPUs have been well explored in prior work, custom hardware designs for efficiently exploiting the algorithm in hardware remain under-explored.
A systolic array design concept for implementing \sa for one level of recursion on 2\by2 matrices has been proposed in the work by Elfimova \ea \cite{elfimova2001fast} without evaluation of an implementation.
Another hardware design for implementing \sa for one level of recursion on 2\by2 matrices has also been proposed in the work by León-Vega \ea \cite{leon2023accel}, where the Strassen architecture reduced FPGA DSP usage by up to 12.5\% at the expense of 25-40\% increase in LUT resources to implement the additional adders.

Unlike the only two prior works on custom hardware designs for executing the Strassen algorithm, we propose architectures in this work that allow for \sa to be implemented on matrices larger than 2\by2.
This is essential for minimizing the complexity penalty of the additional adders.
Additionally, the architectures are capable of implementing multiple levels of \s recursion to achieve greater hardware resource savings.
Furthermore, the proposed architectures allow proven traditional systolic arrays to be still used at the core.
Alternatively, they can allow \sa to be used in combination with other hardware designs that can efficiently perform further algebraic optimizations on matrices after the Strassen portion is carried out, such as techniques from our prior work \cite{pogue2024fast}.
Finally, the proposed Strassen architectures are multisystolic array designs, meaning they can multiply smaller matrices with higher utilization than single-systolic array designs with the same computational strength.

\subsection{Notation}
The following notation is used throughput the remainder this work for describing different systolic array architectures or their workloads:
\begin{itemize}
\item $r$: The number of recursion levels in \eq{smm:eq:mm} or \seq that are implemented in a hardware architecture.
\item \mm?\zero: A traditional single-systolic array implementing conventional matrix multiplication \eq{smm:eq:mmZero} in hardware.
\item \mm?r: A traditional multisystolic array implementing conventional blocked matrix multiplication \eq{smm:eq:mm} in hardware for $r$ levels of recursion.
\item \smm: The proposed Strassen multisystolic array implementing \seq in hardware for $r$ levels of recursion.
\item MXU: In this work, systolic arrays may also be referred to as matrix multiplication units (MXU)s for convenience.
\item (S)MM$_{(r)}$ $X$\by$Y$: An \mm?\zero, \mm?r, or \smm?r architecture may also be referred to with two numbers $X$\by$Y$ specified beside it.
  Here, $X$ and $Y$ represent the width and height, respectively, in number of MAC units of each \mm?\zero systolic array instantiated at the lowest level of recursion in the architecture.
  For example, an \mm?\zero 64\by64 MXU (meaning $X = Y = 64$) would contain $64^2$ MAC units, an \mm?1 32\by32 MXU (meaning $r = 1$ and $X = Y = 32$) would contain $8^1\times32^2$ MAC units, and an \smmArch?2 8\by8 MXU (meaning $r = 2$ and $X = Y = 8$) would contain $7^2\times8^2$ multipliers.

\item $n$: The width/height of the matrices that are being fed as inputs to a systolic array to be multiplied.
\end{itemize}

\section{Strassen Architecture}
\label{smm:sec:arch}
The proposed architectures achieve a more efficient implementation of \sa than what is possible through execution on CPUs and GPUs by pipelining and performing the extra additions and data movement steps at all levels of recursion in parallel with the matrix multiplications.
The architectures are functionally equivalent to conventional multisystolic array designs while allowing the theoretical complexity reductions of \sa to be translated directly into hardware resource savings.

\subsection{Memory Layout and Access Algorithm}
\label{smm:sec:mem-layout}

In order to perform the extra Strassen data movement and addition steps at all levels of recursion in parallel with the matrix multiplications, the architecture reads one row/column at a time of the \A and \B input matrix \subblocks from the lowest level of recursion in \eq{smm:eq:strass-first} simultaneously.
This generates and provides all \T and \Ss \subblocks one row/column at a time for performing all the matrix multiplications in \eq{smm:eq:strass-last} at the lowest level of recursion in parallel.
The \T and \Ss \subblocks are all immediately generated from the \A and \B input \subblocks and consumed in parallel like this to eliminate any additional execution time or hardware resources needed for storing/re-accessing them for later use.

\begin{figure}
  \centering
  \includegraphics[scale=.75]{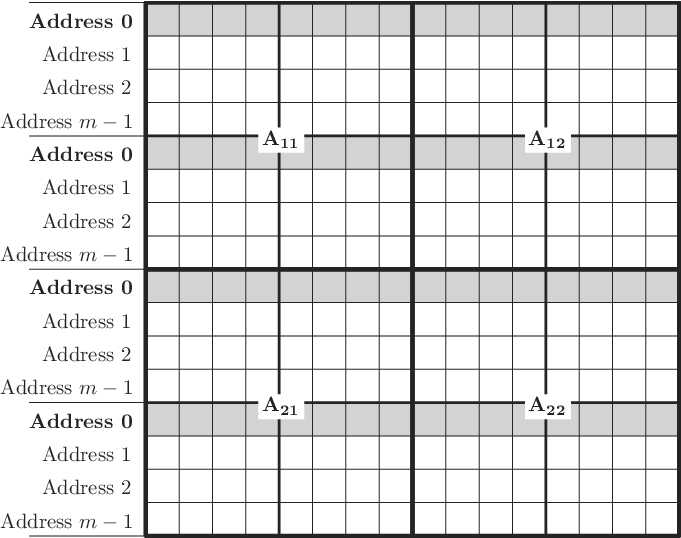}
  \caption{Example data layout for the \A matrix in memory for an architecture implementing Strassen matrix multiplication for 2 levels of recursion  (\smmArch?2).
    Each address $i$ contains every $m^{th}$ row of \A concatenated together starting at row $i$ (notated as \m.A'{i:m:,:}).
    To help illustrate this, the gray coloured rows are all elements of \A belonging to address 0, which forms \m.A'{0:m:,:} containing row 0 of every \A \subblock from the lowest level of recursion in \eq{smm:eq:strass-first}.
    The organization for the \B matrices in memory are the same, except that the order of the elements is transposed compared to the \A matrix layout shown here.}
  \label{smm:fig:mem}
\end{figure}

To achieve this, each \AB matrix fed into the MXU is divided into $4^r$ equal \subblocks of size $m$\by$k$ for \A and of size $k$\by$n$ for \B, where each row/column $i$/$j$ of each \A/\B \subblock is stored in the accelerator's \A and \B memories at location $i$/$j$ plus an offset.
An example of this memory layout for implementing 2 levels of Strassen recursion is shown in \fig{smm:fig:mem}.
This means that each \A memory location $i$ is a vector containing every $m^{th}$ row of \A starting at row $i$ concatenated together (notated as \Av), and each \B memory location $j$ is a vector containing every $n^{th}$ column of \B starting at column $j$ concatenated together (notated as \Bv).
This allows one row or column of all $4^r$ \A/\B \subblocks from the lowest level of recursion in \eq{smm:eq:strass-first} to all be read at once from a single memory location and fed into the MXU each clock cycle.
\Av and \Bv rows/columns are then read consecutively when feeding the \AB blocks into the MXU.

As shown in \eq{smm:eq:strass-first}, the input \AB matrices at each level of recursion are divided into four \block quadrants labelled \An and \Bn of size $M$\by$K$ for \An quadrants and of size $K$\by$N$ for \Bn quadrants.
The portions of each \Av and \Bv vector belonging to quadrant \An and \Bn are notated as \Anv and \Bnv.
The MXU then computes and returns row $i$ of all \C \subblocks from the lowest level of recursion in \eq{smm:eq:strass-last} in every clock cycle $i$, allowing \Cv to be stored in the same format as \A in memory for if \C will later be taken as an \A input for a later matrix multiplication.

\subsection{Strassen Multisystolic Array Design}
\begin{figure}
  \centering
  \includegraphics[scale=.89]{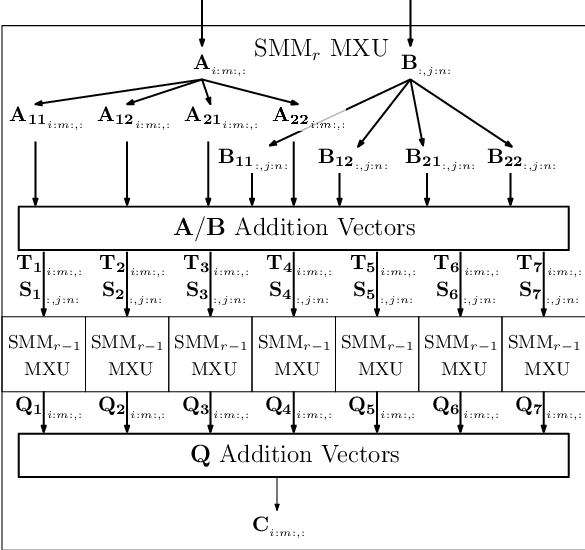}
  \caption{Top-level diagram of the proposed \smmArch multisystolic array architecture for implementing Strassen matrix multiplication \seq for $r$ levels of recursion in hardware.}
  \label{smm:fig:smm-mxu}
\end{figure}
\begin{figure}
  \centering
  \includegraphics[scale=.89]{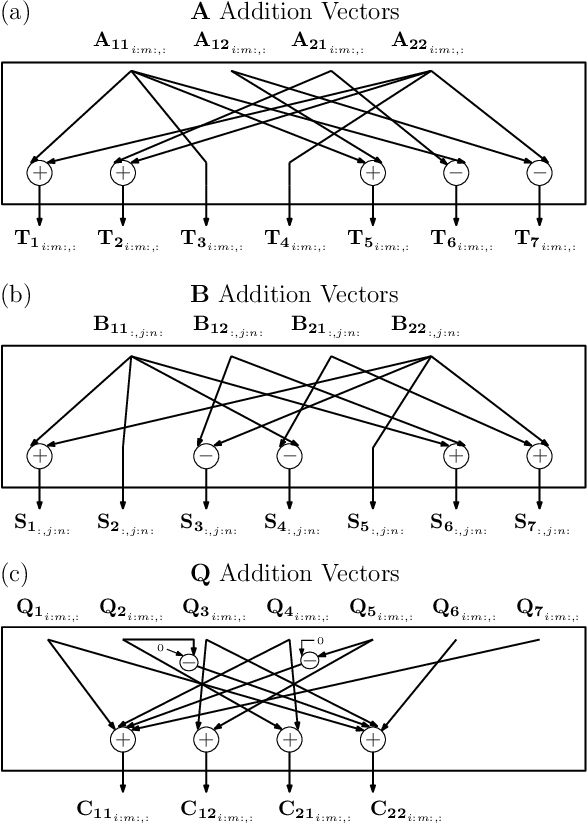}
  \caption{Internal structure of the \smmArch MXU addition vectors from \fig{smm:fig:smm-mxu}.
  }
  \label{smm-add-units}
\end{figure}

\Figure \ref{smm:fig:smm-mxu} shows the proposed \smmArch multisystolic array architecture.
Rather than having one $X$\by$Y$ MXU with $X$ columns and $Y$ rows of MAC units for efficiently multiplying matrices down to size $X$\by$Y$, this architecture consists of $7^r$ smaller $X/2^r$\by$Y/2^r$ MXUs that together efficiently multiply matrices down to the same size but at a higher throughput.
Furthermore, it achieves this with fewer MAC units than a conventional multisystolic array design.
This both allows smaller matrices to be multiplied at a higher utilization and increases the throughput per MAC unit.

The \Av and \Bv vectors read into the MXU are first divided into their four \Anv and \Bnv portions depending on which quadrant of \A/\B each element belongs to as shown in \fig{smm:fig:smm-mxu}.
They then pass through the \A/\B addition vectors shown in \fig{smm-add-units} (a) and (b) to form the \Tv/\Sv matrices.
The \A/\B addition vectors both contain 5 addition vectors each consisting of $K$ scalar adders or subtractors, where $K$ is the width of the four \m.A?{ij} \blocks and the height of the four \m.B?{ji} \blocks as defined in \secn{smm:sec:mem-layout}.
The 7 \Tv/\Sv vectors then pass into the next level of \smmArch?{r-1} MXUs to perform the 7 matrix \block multiplications.
The \Qv vectors of the matrix \block multiplication outputs then pass through the \Qv addition vectors shown in \fig{smm-add-units} (c) consisting of 8 addition vectors each containing $N$ scalar adders or subtractors.
This forms the final \C product, where $N$ is the width of the four \m.B?{ji} \blocks as defined in \secn{smm:sec:mem-layout}.

Each of the 7 \smmArch?{r-1} MXUs can contain 7 more \smmArch?{r-2} MXUs for implementing another level of \s recursion and repeating the process above, or they can be instantiated as a baseline \mmArch?\zero MXU shown in \Figure \ref{smm:fig:MM-mxu}.
For implementing the next level of \smmArch?{r-2} MXUs inside each \smmArch?{r-1} MXU, each \Tv/\Sv input passed into an \smmArch?{r-1} MXU will then be considered as the full \Av/\Bv inputs within that MXU and are split again into the next level of four \Anv/\Bnv vectors.
The dimensions of the matrix \blocks being read/computed and the number of scalar adders in the addition vectors within each \smmArch?{r-1} MXU will then be reduced by a factor of 2 at each level of recursion.
For fixed-point implementations, the \Tv/\Sv inputs to each \smmArch?{r-1} MXU that were formed from an addition or subtraction in the \A or \B vector addition units will have an increased bitwidth by 1 bit.

\subsection{Baseline Designs}

\begin{figure}
  \centering
  \includegraphics[scale=1.2]{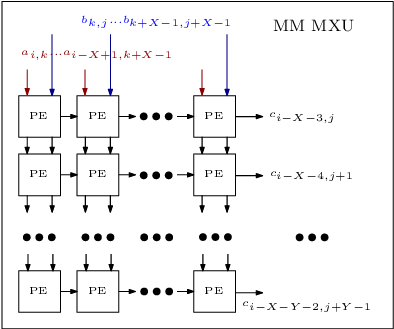}
  \caption{Baseline \mmArch?\zero single-systolic array architecture that implements conventional matrix multiplication \eq{smm:eq:mmZero} in hardware, provided for completeness and clarity.
    It is instantiated at the lowest level of recursion in the \smmArch and \mmArch MXU architectures.
    $X$ here represents the width of the $a$ and $b$ vectors entering the \mmArch?\zero MXU, and $Y$ represents the width of the $c$ vectors exiting the MXU.}
  \label{smm:fig:MM-mxu}
\end{figure}

\begin{figure}
  \centering
  \includegraphics[scale=.78]{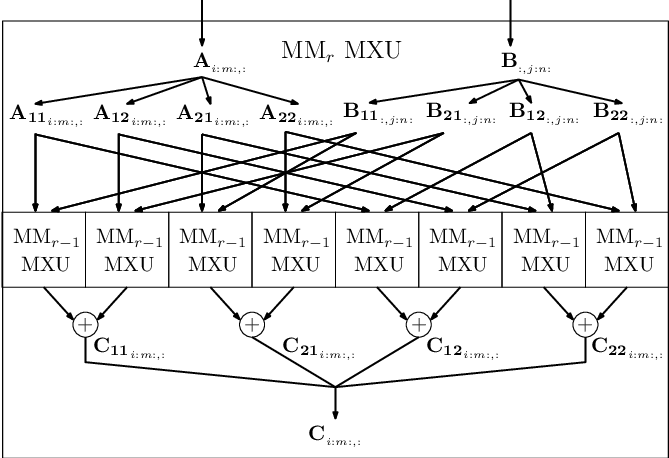}
  \caption{Baseline \sAlg.{\mm?r} multisystolic array architecture for implementing conventional blocked matrix multiplication \eq{smm:eq:mm} for $r$ levels of recursion in hardware.}
  \label{smm:fig:multi-mm-mxu}
\end{figure}

\begin{figure}
  \centering
  \includegraphics[scale=.8]{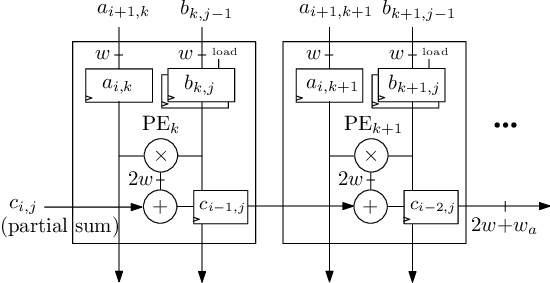}
  \caption{The internal PE structure of each \mmArch?\zero MXU from \fig{smm:fig:MM-mxu}, provided for completeness.
    Here, $w_a$ is the additional bitwidth added to account for accumulation, equal to $\ceil{\tx{log}_2(X)}$, where $X$ is the width of the $a$ and $b$ vectors entering the \mmArch?\zero MXU.
  }
  \label{smm:fig:pes}
\end{figure}

We later compare the \smmArch?r architectures with baseline \mmArch multisystolic array architectures shown in \fig{smm:fig:multi-mm-mxu} which execute \eq{smm:eq:mm} in parallel for $r$ levels of recursion.
The baseline \mmArch architectures are functionally identical to the \smmArch architectures, but they consist of $8^r$ smaller $X/2^r$\by$Y/2^r$ MXUs rather than $7^r$.
\Figure \ref{smm:fig:MM-mxu} also shows the internal structure of each baseline \mmArch?\zero MXU present at the lowest level of recursion in each \smmArch and \mmArch architecture, and \fig{smm:fig:pes} shows the internal structure of the processing elements (PE)s inside the \mmArch?\zero MXUs.

\section{Results}

In this section, we evaluate example implementations of the proposed \smm architectures.
In \secn{smm:sec:results-baseline}, we compare the \smm MXU architectures in isolation against our baseline MXU designs in Table \ref{smm:tab:smm-MM}.
In \secn{smm:sec:results-prior-work} we evaluate the \smm MXU architectures in Tables \ref{smm:tab:second}-\ref{smm:tab:last} compared to prior work when integrated into an end-to-end deep learning accelerator system based on the system from our previous work \cite{pogue2024fast}, which has open-source code \cite{ffip-source}.
We first describe the integration of our proposed systolic arrays into the deep learning system in \secn{sec:system}, and in Sections \ref{sec:mu} and \ref{smm:sec:mat-sz}, we define performance metrics used to compare the \smmArch architectures against baseline designs and prior works.

\subsection{System Integration}
\label{sec:system}
We were able to integrate the \smm architectures into a system based on our previous work \cite{pogue2024fast}, which has open-source code \cite{ffip-source}, by swapping the \smmArch MXU architectures from \fg{smm:fig:smm-mxu} into our system design \cite{pogue2024fast} in place of the free-pipeline fast inner-product (FFIP) MXU.

In order to perform GEMM on the proposed MXUs and multiply matrices of arbitrary sizes that can be larger than the MXU dimensions, the full \AB matrices are first divided into \gemmtiles prior to being divided further into smaller \blocks for executing \eq{smm:eq:mm} or \seq.
The \gemmtiles are then fed into the MXU one-by-one.
Each \gemmtile is then considered as the full \AB matrix from \eq{smm:eq:mm} or \seq while being fed into the MXU and gets further divided into smaller \m.A?{ij}/\m.B?{ji} \blocks within the MXU.

Following each \gemmtile multiplication, the partial \gemmtile products are accumulated outside of the MXU to generate each final \gemmtile product.
Prior to each \gemmtile multiplication, a $\bld{B}$ \gemmtile is loaded into the MXU.
It then remains in place as the $\bld{A}$ \gemmtile flows through the MXU producing the \gemmtile product, during which a new \Av vector is fed into the MXU each clock cycle.
Additionally, to hide the latency of loading $\bld{B}$ \gemmtiles, the MXU PEs each contain one extra $b$ buffer to load the next $\bld{B}$ \gemmtile into the MXU as the current \gemmtile is being multiplied.

Each \A, \B, and \C \subblock entering or exiting the top-level MXU for the \smm and baseline MXUs first pass through triangular-shaped register arrays each containing $X$ shift registers of varying depths.
Here, each shift register $SR_k$ has a depth of $k$ and loads one $a_{i,k}$ or $b_{k,j}$ element per clock cycle.
These triangular buffers are explained further in our prior work \cite{pogue2024fast} and they allow the vector elements to enter the MXU in the necessary order as depicted in the element indices in \figs \ref{smm:fig:MM-mxu} and \ref{smm:fig:pes}.

\subsection{Multiplier Compute Efficiency}
\label{sec:mu}
In this subsection, we define an efficiency metric called the multiplier compute efficiency (MCE) in \eq{smm:eq:mu} which we use to compare the \smmArch architectures against baseline designs and prior works.
This is used to quantify how much the algebraic optimizations exploited in an architecture reduce the computational complexity.
Reductions in computational complexity allow an architecture to utilize its multipliers more effectively than conventional designs using no algebraic optimizations.
The multiplier compute efficiency is defined as follows:
\begin{align}  \label{smm:eq:mu}
  \tx{MCE} = \frac{\tx{mults} / \tx{multiplier}} {\tx{clock cycle}}
  &= \frac{(\tx{mults/s})/\tx{\#multipliers}}{f}
  \,.
\end{align}
Here, mults/s above is measured by taking the number of multiplications required to carry out an execution using conventional algebra and dividing it by the measured execution time.
Finally, \#multipliers is the number of instantiated multipliers in the design, and $f$ is the clock frequency that the hardware design is operating at.

Conventional matrix multiplication algorithms such as \eq{smm:eq:mm} have no algebraic optimizations for reducing the computational complexity.
Therefore, the limit/maximum achievable value (also referred to as the roof) of the metric in \eq{smm:eq:mu} is the following when using conventional matrix multiplication in hardware:
\begin{align}  \label{smm:eq:mm-mu-roof}
  \tx{roof}\lb\tx{MCE}_{\tx{\mm?r}}\rb &= 1
  \,.
\end{align}
In contrast, \sa requires $8^r/7^r$ times fewer multiplications than a conventional matrix multiplication algorithm, where $r$ is the number of levels of recursion implemented in \sa.
Therefore, the multiplier compute efficiency can reach the following limit in \smmArch architectures:
\begin{align} \label{smm:eq:mu-roof}
  \tx{roof}\lb\tx{MCE}_{\tx{\smm?r}}\rb &= 
  \lb\frac{8}{7}\rb^{r}
  \,.
\end{align}

As discussed in \secn{smm:sec:background}, \sa reduces the overall number of operations in matrix multiplication.
Furthermore, any additions required before the matrix multiplications in the algorithm are even less of a concern in fixed-point implementations.
This is because the hardware complexity of fixed-point multipliers typically scale quadratically with the input bitwidth compared to linearly for adders and registers \multComplexity, causing the hardware footprint of multipliers to dominate that of adders and registers.

However, one of the impediments of using \sa for fixed-point implementations is that the bitwidths of the multiplication inputs increase by $r$ bits for $r$ levels of Strassen recursion that are implemented, reducing its potential area savings for custom fixed-point hardware designs.
Nonetheless, this impediment for fixed-point designs can be inherently mitigated in FPGA implementations so long as $r$ plus the initial input width is not larger than the maximum input width supported by the FPGA's DSP units.
For example, each DSP in common Intel/Altera FPGAs instantiate two 18\by19-bit multipliers \cite{intel-dsp}, and common input bitwidths for applications such as deep learning are 16 bits or less.
This leaves room for at least 2 or more levels of Strassen recursion to be implemented before surpassing the bitwidth limit supported by the DSPs.

Furthermore, due to the flexible nature of custom hardware design, the \smm?r architectures can be efficiently mapped onto other DSP units in general which support input bitwidths up to $n$ bits by customizing the input datapath bitwidth $w$ and value of $r$ as necessary to ensure that $w+r \le n$.
So long as the accuracy requirements of the application are still met, this will allow the \smm?r designs and their increase in multiplier bitwidth to still be efficiently mapped onto DSP units of any bitwidth in a general way.

\subsection{Supporting Smaller Matrices with the Same Performance}
\label{smm:sec:mat-sz}
Multisystolic array designs such as the \smmArch and baseline \mm?r architectures have the ability to efficiently multiply smaller matrices than a single-systolic array design with the same performance capability.
By executing \eq{smm:eq:mm} or \seq fully in parallel for $r$ levels of recursion, matrix products of size as small as $n\times n$ can be computed up to once every $n/2^r$ clock cycles in an \mm?r or \smmArch multisystolic array design.
Furthermore, these matrix products require $n^3$ multiplications to calculate using conventional algebra.
Therefore, the ratio of an architecture's throughput per clock cycle versus its smallest supported matrix sizes it can multiply, which we refer to as the matrix size efficiency (MSE), is the following:
\begin{align} \label{smm:eq:mat-sz}
  \tx{MSE} = 
  \frac{\tx{mults}/\tx{clock cycle}} {\tx{min. mat. size (h\by w)}}
  \,,
\end{align}
which has the following roof for multisystolic arrays:
\begin{align} \label{smm:eq:mat-sz}
  \tx{roof}\lb\tx{MSE}_{\tx{(S)MM}_r}\rb
  = \frac {n^3/\lb n/2^r\rb} {n\times n} = 2^r
  \,.
\end{align}

In contrast, a single-systolic array design can produce matrix products of size as small as $n\times n$ up to once every $n$ clock cycles, making this ratio the following for a single-systolic array design:
\begin{align}
  \tx{roof}\lb\tx{MSE}_{\tx{MM}}\rb
  = \frac {n^3/ n} {n\times n} = 1
  \,.
\end{align}
This shows that the \smmArch and baseline \mm?r multisystolic array designs can efficiently multiply matrices $2^r$ times smaller than a single-systolic array architecture with the same performance capability.

As discussed in \secn{multi-sys}, this is an important property for increasing a systolic array accelerator's maximum achievable throughput on real-life workloads.
Even if more compute resources are instantiated to scale up the size of the systolic array, the systolic array will begin to be underutilized after its size surpasses the workload's matrix sizes.
This is particularly true in modern workloads such as deep learning acceleration, where the matrix sizes that the workloads break down to can be smaller than the maximum systolic array size that could be instantiated in an accelerator \citeMultiSys.
In \secn{smm:sec:results-prior-work}, we demonstrate how this property allowed us to scale up our deep learning accelerator design without compromising utilization to achieve state-of-the-art ResNet \cite{kaiming2016deep} throughput.

\subsection{Comparison to Baseline Designs}
\label{smm:sec:results-baseline}
\newcolumntype{D}{>{\centering\arraybackslash}p{1.98cm}}
\newcolumntype{E}{>{\centering\arraybackslash}p{1.8cm}}
\begin{table*}\centering
\caption{Comparing \smmArch multisystolic array architectures against the baseline \mmArch?\zero single-systolic array architecture and baseline \mmArch multisystolic array architectures in isolation (without integration into a deep learning accelerator system).}
\label{smm:tab:smm-MM}
\label{smm:tab:first}
\scriptsize
\begin{threeparttable}
  \begin{tabular}{|>{\raggedleft}p{3.3cm}|D|DD|DDE|}\toprule
\arrayrulecolor{black}
&\alg.\MM?\zero<{} 48\by48 &\alg.\MM?1<{} 16\by16  &\alg.\SMM?1<{} 16\by16 &\alg.\MM?2<{} 6\by6 &\alg.\SMM?2<{} 6\by6 &\alg.\SMM?2<{} 6\by6 (with extra regs.)   \\
\toprule
DSPs                                 &1,152                     &1,024                 &\textbf{896}           &1,152               &\textbf{882}         &\textbf{882}            \\
\arrayrulecolor{black!30}\midrule
\arrayrulecolor{black}
ALMs                                 &34,890                    &30,872                &30,265                 &36,397              &35,863               &38,485                  \\
\arrayrulecolor{black!30}\midrule
\arrayrulecolor{black}
Registers                            &130,262                   &118,049               &115,830                &138,219             &133,511              &147,750                 \\
\arrayrulecolor{black!30}\midrule
\arrayrulecolor{black}
Frequency (MHz)                      &399                       &398                   &380                    &388                 &291                  &361                     \\
\arrayrulecolor{black}\midrule
$\tx{roof}\lb\tx{Throughput}\rb$ (GOPS) \tnote{1}     &1839     &1630                  &1556                   &1788                &1341                 &1663                    \\
\arrayrulecolor{black!30}\midrule
Throughput/DSP \tnote{2}                       &1.60            &1.59                  &\textbf{1.74}                   &1.55                &1.52                 &\textbf{1.89}                    \\
\arrayrulecolor{black!30}\midrule
\macUtA \tnote{3}                              &1               &1                     &\textbf{1.14}          &1                   &\textbf{1.31}        &\textbf{1.31}           \\
\arrayrulecolor{black}\midrule
Min. supported matrix size \tnote{4} &48\by48                   &32\by32               &32\by32                &24\by24             &24\by24              &24\by24                 \\
\arrayrulecolor{black!30}\midrule
\minSzMetric \tnote{5}               &1                         &2                     &2                      &4                   &4                    &4                       \\
\arrayrulecolor{black}
\bottomrule
\end{tabular}
  \begin{tablenotes}
    \item All designs are synthesized on Arria 10 GX 1150 FPGA for 16-bit fixed-point inputs and consume 0 memory resources.
\item[1] Maximum achievable throughput in giga operations per second, where throughput is equal to the number of operations required to carry out an execution using conventional algebra divided by the measured execution time.
\item[2] Shows which designs can achieve the highest throughput for the same number of DSPs.
\item[3] Maximum achievable multiplier compute efficiency, defined in \secn{sec:mu}, measures how effectively the architecture can utilize its multipliers. It can surpass 1 in \smm architectures because the observed mults/s is equal to the number of multiplications required to carry out an execution using conventional algebra divided by execution time.
\item[4] Minimum input matrix sizes that can be multiplied at peak throughput/full utilization.
\item[5] \matSzExpl
\end{tablenotes}
\end{threeparttable}
\end{table*}

Table \ref{smm:tab:smm-MM} shows the resource usage and performance comparison between the proposed \smmArch and baseline \mm{/}\mmArch systolic array architectures in isolation (without integration into a deep learning accelerator system).
The \smmArch?1 and \smmArch?2 architectures overall have a similar amount of soft logic resources and the same throughput per clock cycle roof as the \mmArch?1 and \mmArch?2 architectures, respectively, but they require \textbf{1.14-1.31\x fewer DSP units}.
Compared to the multisystolic array \mm?1 and \mm?2 designs, the \smmArch?1 and \smmArch?2 architectures are also functionally equivalent, respectively, other than having a lower clock frequency.
To help mitigate the limitation of having a lower frequency, we added an extra \smmArch?2 design (which had the biggest issue with clock frequency) on the far right of Table \ref{smm:tab:smm-MM} containing additional pipelining registers in the addition logic of each Q Addition Vectors unit from \fig{smm-add-units} (c).
This extra design demonstrates how a trade-off can be optionally made to increase the design's clock frequency at the cost of some extra soft logic resources.

Nonetheless, the lower clock frequencies of the \smm?r designs in Table  \ref{smm:tab:smm-MM} are compensated by the fact that the \smm?r designs achieve more effective operations from the same number DSP units.
Since the reduction in DSP units is greater than the reduction in clock frequency in the \smm?1 design and \smm?2 design with extra registers relative to their \mm?r counterparts, they would be able to achieve a higher overall throughput if scaled up in size to use the same number of DSPs.
This is shown by the Throughput/DSP metric in Table  \ref{smm:tab:smm-MM}, which shows that the \smm?r designs achieve up to \textbf{22\% more throughput per DSP} than their \mm?r counterparts.
Finally, if the frequency-limiting critical path is in external control or other logic outside of the systolic array after integrating it into an end-to-end accelerator system, as was the case in our full-system accelerators from Tables \ref{smm:tab:8}-\ref{smm:tab:ffip}, this limitation of a lower frequency is further mitigated.

\begin{figure}[]
  \hspace*{-0.25cm}
  \centering
  \includegraphics[scale=0.0525]{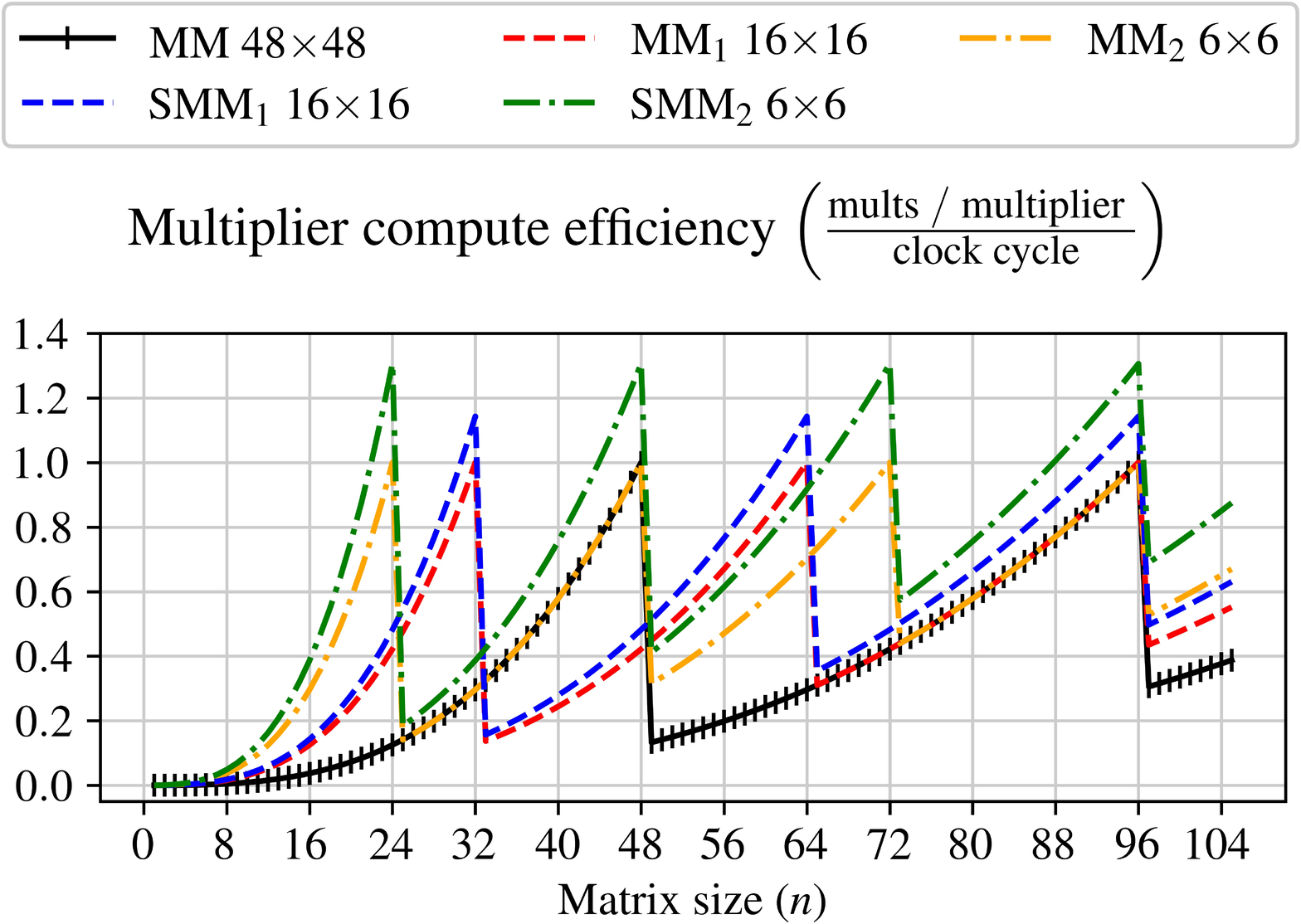}
  \vspace*{-0.6cm}
  \caption{Plotting the multiplier compute efficiency of the architectures in Table \ref{smm:tab:smm-MM} when multiplying different input matrices of size $n$\by$n$.
    As shown, the minimum matrix size that can be efficiently multiplied gets smaller in the order of the \mm?\zero{,} \mm?1{/}\smm?1{,} and \mm?2{/}\smm?2 designs, respectively.}
  \label{smm:fig:perf}
\end{figure}

The throughput per clock cycle roof of the \mm?\zero and \mm?2 baseline designs in Table \ref{smm:tab:smm-MM} are equal and they consume the same number of DSP resources, but the \mm?\zero design requires slightly fewer ALM and register resources.
However, this penalty may be justified in the \mm?2 design when considering that the minimum matrix size (height\by width) that can be multiplied while fully utilizing the MXU is 4$\times$ smaller in the \mm?2 design compared to the \mm?\zero design.
This increases its performance scalability for accelerating modern workloads such as deep learning as discussed in \secn{multi-sys} and \ref{smm:sec:mat-sz}.
This ability of the multisystolic array designs to more efficiently multiply smaller matrices is further illustrated in \fig{smm:fig:perf}.
This same property is true for the \smmArch?2 design, except it achieves this with fewer DSP resources.
This benefit is demonstrated in \secn{smm:sec:results-prior-work}, where this property allowed us to scale up our deep learning accelerator design without compromising utilization to achieve state-of-the-art ResNet throughput.

\subsection{Comparison to Prior Work}
\label{smm:sec:results-prior-work}

Full system-level validation of the experimental accelerator as integrated into the system from our previous work \cite{pogue2024fast} has been done on an Arria 10 SoC Developement Kit \cite{sx-dev-kit} containing the \sx device by measuring throughput in real-time.
However, this device contains fewer soft logic resources than the \gx used in the prior works we compare against, and we generate compilation results for our design on the same \gx device used in prior works for a more fair and consistent comparison.
Throughput values of our designs on the \gx device are then calculated using an accurate throughput estimation model based on our highly deterministic and time-predictable system implementation, which accurately predicts actual throughputs measured on the \sx device available to us.
Tables \ref{smm:tab:second}-\ref{smm:tab:last} show throughputs for ResNet \cite{kaiming2016deep} neural networks.

The works from Liu \ea \cite{liu2022toward} and Fan \ea \cite{fan2022fpga} in Table \ref{smm:tab:8} use a technique to pack two 8-bit multiplications onto each 18\by19-bit multiplier in the DSPs and additional ALMs, and therefore the number of multipliers is calculated as \#DSP\by4 in those works.
The number of multipliers in the works \cite{huang2022fpga}, \cite{kim2023agamotto} from Table \ref{smm:tab-mixed} is calculated as \#DSPs since they are implemented on AMD FPGAs where each DSP instantiates one 18\by27-bit multiplier \cite{amd-dsp}.
In Tables \ref{smm:tab:8} and \ref{smm:tab-mixed}, the number of multipliers in the prior works \cite{an2022opencl}, \cite{dai2024dcp} is equal to \#DSPs\by2, where each DSP in the Altera FPGAs instantiates two 18\by19-bit multipliers \cite{intel-dsp}.
The number of multipliers used in the MXUs from our architectures in Tables \ref{smm:tab:first}-\ref{smm:tab:second} is equal to $8^r$ or $7^r$ times $X$\by$Y$ for the \mmArch and \smmArch MXUs, respectively.
For example, an \mm?\zero 64\by64 MXU (meaning $r = 0$ and $X = Y = 64$) would contain $8^0\times64^2$ multipliers, an \mm?1 32\by32 MXU would contain $8^1\times32^2$ multipliers, and an \smmArch?2 8\by8 MXU would contain $7^2\times8^2$ multipliers.
Due to the FFIP reduction in multipliers as described in our prior work \cite{pogue2024fast}, the number of multipliers for the FFIP architectures in Table \ref{smm:tab:ffip} is equal to $8^r$ or $7^r$ times $X\times Y/2 + X/2$ for the FFIP and FFIP+\smmArch designs, respectively.
Additionally, for our deep learning accelerator implementations in Tables \ref{smm:tab:second}-\ref{smm:tab:ffip}, there are an additional $Y\times4^r$ multipliers located outside the MXU in the Post-GEMM Unit \cite{pogue2024fast} for performing inter-layer quantization rescaling functions.
For our designs requiring more than 3036 multipliers, 3036 are instantiated on 1518 DSPs, and the remainder are instantiated in soft logic resources as the DSP resources are fully utilized.

\label{smm:sec:results-prior-work}
\begin{table*}\centering
  \caption{\smmArch multisystolic array architectures integrated into a deep learning accelerator system compared with prior state-of-the-art deep learning accelerators.}
\label{smm:tab:8}
\label{smm:tab:second}
\scriptsize
\begin{threeparttable}
  \begin{tabular}{|>{\raggedleft}p{2.2cm}|AA|AA|AA||AAA|AAA|}\toprule
    \arrayrulecolor{black}
    &\multicolumn{2}{c|}{TNNLS '22 \cite{liu2022toward}} &\multicolumn{2}{c|}{TCAD '22 \cite{fan2022fpga}} &\multicolumn{2}{c||}{Entropy '22 \cite{an2022opencl}} &\multicolumn{3}{c|}{\alg.\SMM?1<{} 32\by32} &\multicolumn{3}{c|}{\alg.\SMM?2<{} 8\by8}                                                           \\
\toprule
DSPs                               &\multicolumn{2}{c|}{1473}                           &\multicolumn{2}{c|}{1473}                        &\multicolumn{2}{c||}{1503}                           &\multicolumn{3}{c|}{1518}               &\multicolumn{3}{c|}{1518}              \\
\arrayrulecolor{black!30}\midrule
\arrayrulecolor{black}
ALMs                               &\multicolumn{2}{c|}{304K}                           &\multicolumn{2}{c|}{304K}                        &\multicolumn{2}{c||}{303K}                           &\multicolumn{3}{c|}{306K}               &\multicolumn{3}{c|}{145K}              \\
\arrayrulecolor{black!30}\midrule
\arrayrulecolor{black}
Registers                          &\multicolumn{2}{c|}{889K}                           &\multicolumn{2}{c|}{890K}                        &\multicolumn{2}{c||}{-}                              &\multicolumn{3}{c|}{641K}               &\multicolumn{3}{c|}{386K}              \\
\arrayrulecolor{black!30}\midrule
\arrayrulecolor{black}
Memories                           &\multicolumn{2}{c|}{2334}                           &\multicolumn{2}{c|}{2334}                        &\multicolumn{2}{c||}{1953}                           &\multicolumn{3}{c|}{2713}               &\multicolumn{3}{c|}{2036}              \\
\arrayrulecolor{black!30}\midrule
\arrayrulecolor{black}
Frequency (MHz)                    &\multicolumn{2}{c|}{200}                            &\multicolumn{2}{c|}{220}                         &\multicolumn{2}{c||}{172}                            &\multicolumn{3}{c|}{293}                &\multicolumn{3}{c|}{295}               \\
\arrayrulecolor{black!30}\midrule
\arrayrulecolor{black}
Model                              &ResNet-50      &VGG 16                               &Bayes ResNet-18  &Bayes VGG 11                    &R-CNN (ResNet-50) &R-CNN (VGG 16)                     &ResNet-50 &ResNet-101 &ResNet-152       &ResNet-50 &ResNet-101 &ResNet-152                  \\
\arrayrulecolor{black}\midrule
\tabThpt                           &1519           &1295                                &1590             &534                            &719               &865                               &3750      &4116       &4276             &2024      &2115       &2158                       \\
\arrayrulecolor{black!30}\midrule
\arrayrulecolor{black}
\arrayrulecolor{black}
\smmMacUt                             &0.645          &0.550                               &0.639            &0.206                          &0.696             &0.837                             &0.877     &0.963      &1.002            &1.051     &1.098      &1.120         \\
\arrayrulecolor{black}
\bottomrule
\end{tabular}
\begin{tablenotes}
\item All designs are synthesized on Arria 10 GX 1150 FPGA for 8-bit fixed-point inputs.
\item[1] Throughput in giga operations per second, equal to the number of operations required to carry out an execution using conventional algebra divided by execution time.
\item[2] \smmMacUtExpl
\end{tablenotes}
\end{threeparttable}
\end{table*}

\begin{table*}\centering
  \caption{Comparison of an FFIP \cite{pogue2024fast} single-systolic array architecture, which doubles performance per MAC unit, with combined FFIP+\smmArch multisystolic array architectures when integrated into deep learning accelerator systems.}
\label{smm:tab:ffip}
\label{smm:tab:last}
\scriptsize
\begin{threeparttable}
  \begin{tabular}{|>{\raggedleft}p{2.2cm}|AAA||AAA|AAA|}\toprule
\arrayrulecolor{black}
                                   &\multicolumn{3}{c||}{TC '24 \cite{pogue2024fast} (FFIP 64\by64)}            &\multicolumn{3}{c|}{\alg.\sffip?1<{} 32\by32}              &\multicolumn{3}{c|}{\alg.\sffip?2<{} 8\by8}                                              \\
\toprule
DSPs                               &\multicolumn{3}{c||}{\eDSPs}                                                &\multicolumn{3}{c|}{1518}                               &\multicolumn{3}{c|}{946}                                                            \\
\arrayrulecolor{black!30}\midrule
\arrayrulecolor{black}
ALMs                               &\multicolumn{3}{c||}{\eALMs}                                                &\multicolumn{3}{c|}{216K}                               &\multicolumn{3}{c|}{165K}                                                            \\
\arrayrulecolor{black!30}\midrule
\arrayrulecolor{black}
Registers                          &\multicolumn{3}{c||}{\eRegs}                                                &\multicolumn{3}{c|}{627K}                               &\multicolumn{3}{c|}{463K}                                                         \\
\arrayrulecolor{black!30}\midrule
\arrayrulecolor{black}
Memories                           &\multicolumn{3}{c||}{\eMems}                                                &\multicolumn{3}{c|}{2713}                               &\multicolumn{3}{c|}{2036}                                                            \\
\arrayrulecolor{black!30}\midrule
\arrayrulecolor{black}
Frequency (MHz)                    &\multicolumn{3}{c||}{\eFreq}                                                &\multicolumn{3}{c|}{313}                                &\multicolumn{3}{c|}{297}                                                           \\
\arrayrulecolor{black!30}\midrule
\arrayrulecolor{black}
Model                              &ResNet-50            &ResNet-101          &ResNet-152                      &ResNet-50       &ResNet-101     &ResNet-152             &ResNet-50       &ResNet-101     &ResNet-152              \\
\arrayrulecolor{black}\midrule
\tabThpt                           &\eResNetAGOPS        &\eResNetBGOPS  &\eResNetCGOPS                        &4006            &4397           &4568                   &2038            &2130           &2172            \\
\arrayrulecolor{black!30}\midrule
\smmMacUt                             &1.521  &1.655 &1.707                                                   &1.674  &1.837 &1.908            &1.813  &1.895 &1.933 \\
\arrayrulecolor{black}
\bottomrule
\end{tabular}
  \begin{tablenotes}
  \item All designs are synthesized on Arria 10 GX 1150 FPGA for 8-bit fixed-point inputs.
  \footNoteRefs
\end{tablenotes}
\end{threeparttable}
\end{table*}

\newcolumntype{V}{>{\centering\arraybackslash}p{1.4cm}}
\newcolumntype{W}{>{\centering\arraybackslash}p{1.2cm}}
\begin{table}[]\centering
\caption{State-of-the-art deep learning accelerators implemented on other FPGA families for similar neural networks and input bitwidths to provide further comparison with Tables \ref{smm:tab:8} and \ref{smm:tab:ffip}.}
\label{smm:tab-mixed}
\scriptsize
\begin{threeparttable}
  \begin{tabular}{|>{\raggedleft}p{2.5cm}|V|V|W|}\toprule
    \arrayrulecolor{black}
                                          &TNNLS '22 \cite{huang2022fpga} &TCAS-I '23 \cite{kim2023agamotto} &TCAD '24 \cite{dai2024dcp} \\
\toprule
FPGA                                       &AMD VX980                         &AMD XCV U9P                            &Altera Stratix 10 GX650                                              \\
\arrayrulecolor{black!30}\midrule
\arrayrulecolor{black}
Fixed-point input bitwidth                 &8/16  \tnote{3}               &8                                  &8                                                     \\
\arrayrulecolor{black!30}\midrule
\arrayrulecolor{black}
DSPs                                       &3121                          &2048                               &1024                                                  \\
\arrayrulecolor{black!30}\midrule
\arrayrulecolor{black}
ALMs (Altera) / LUTs (AMD)                 &480K                          &-                                  &152K                                                  \\
\arrayrulecolor{black!30}\midrule
\arrayrulecolor{black}
Registers                                  &-                             &-                                  &567K                                                     \\
\arrayrulecolor{black!30}\midrule
\arrayrulecolor{black}
Memories  (20Kb Altera) / (36Kb AMD)    &1457                          &-                                     &2056                                                  \\
\arrayrulecolor{black!30}\midrule
\arrayrulecolor{black}
Frequency (MHz)                            &100                           &200                                &200                                                   \\
\arrayrulecolor{black!30}\midrule
\arrayrulecolor{black}
Model                                      &ResNet-101                    &ResNet-50                          &ResNet-152                                            \\
\arrayrulecolor{black}\midrule
\tabThpt                                   &600                           &287                                &794                                                   \\
\arrayrulecolor{black!30}\midrule
\arrayrulecolor{black}
\smmMacUt                                  &0.961                         &0.351                              &0.969                                                 \\
\arrayrulecolor{black}
\bottomrule
\end{tabular}
\begin{tablenotes}
  \footNoteRefs
\item[3] Weights are quantized to 8 bits and layer input/output is quantized to 8 or 16 bits at different stages.
\end{tablenotes}
\end{threeparttable}
\end{table}

Tables \ref{smm:tab:8}-\ref{smm:tab:ffip} show the \smmArch architectures  integrated into the deep learning system from our previous work \cite{pogue2024fast} compared to state-of-the-art accelerators evaluated on the same FPGA family for the same input bitwidths and similar neural network models.
Integrating the \smmArch multisystolic array design into our deep learning accelerator allowed us to increase the multiplier compute efficiency while also scaling up the computational resources and throughput roof without increasing the minimum supported matrix sizes.
This allowed it to significantly surpass the throughput in our prior work \cite{pogue2024fast} and other state-of-the-art prior works evaluated on the same FPGA family as shown in Tables \ref{smm:tab:8}-\ref{smm:tab:ffip}.
If the design is scaled up using a single-systolic array, the minimum supported matrix size increases and compute resources begin to be underutilized for ResNet execution based on the smaller matrix sizes that its workload decomposes to.
This causes the effective throughput to not increase well despite the design having a larger throughput roof.

The \smmArch?1 32\by32 and \sAlg.\sffip<{}?1 32\by32 designs consume noticeably more memory resources than the \smmArch?2 8\by8 and \sAlg.\sffip<{}?2 8\by8 designs.
However, it is worth noting that this is not due to increased memory requirements, but rather is due to the compiler favouring to swap some register resources for memory resources.
This is because the \smmArch?1 32\by32 and \sAlg.\sffip<{}?1 32\by32 designs have a higher register (and overall area) overhead than the \smmArch?2 8\by8 and \sAlg.\sffip<{}?2 8\by8 designs in order to achieve higher throughput roofs.

In Table \ref{smm:tab:8}, the \smmArch architectures achieve the highest throughput and multiplier compute efficiency compared to the prior works.
The \smmArch?1 and \smmArch?2 architectures' multiplier compute efficiencies in Table \ref{smm:tab:8} approach their limits of 1.14 and 1.31 that are derived in \eq{smm:eq:mu-roof}.
This surpasses the limit of 1 of the baseline \mmArch architectures and prior works that is derived in \eq{smm:eq:mm-mu-roof}, validating \smm's ability to increase multiplier compute efficiency and reduce computational complexity as expected from our analysis.

Table \ref{smm:tab:ffip} shows an example of how \smm can be combined with other algebraic techniques to further increase multiplier compute efficiency limits.
FFIP \cite{pogue2024fast} provides a way to reduce the number
of required multiplications by up to a factor of 2, trading half the multiplications for cheap low-bitwidth additions.
Because of this, the limit for the multiplier compute efficiency metric in \eq{smm:eq:mu} for an FFIP architecture becomes 2, and $2\times(8/7)^r$ for a combined FFIP+\smmArch architecture.
In Table \ref{smm:tab:ffip}, we evaluate architectures that combine FFIP+\smmArch by instantiating \smmArch MXUs that use FFIP MXUs at their lowest level of recursion instead of the conventional \mm?\zero MXUs from \fig{smm:fig:MM-mxu}.
This further increases multiplier compute efficiency compared to a standalone \smmArch or standalone FFIP MXU as seen in the achieved multiplier compute efficiencies of the \sAlg.\sffip<{}?{r} architectures listed in Table \ref{smm:tab:ffip}.

\section{Conclusion}
Strassen's fast matrix multiplication algorithm reduces the complexity of naive matrix multiplication, however, general-purpose hardware is not suitable for achieving the algorithm's promised theoretical speedups.
Furthermore, there is limited prior work on custom hardware architectures designed specifically for executing the algorithm in hardware.
We address this by presenting custom Strassen multisystolic array hardware architectures that are functionally equivalent to conventional multisystolic array designs.
However, they allow the theoretical complexity reductions of \sa to be translated directly into hardware resource savings, even for multiplication of small matrices.

Compared to a conventional multisystolic array design, the proposed architectures implemented on FPGA for 1 and 2 levels of Strassen recursion use 1.14\x and 1.31\x fewer DSP units and an overall comparable amount of soft logic resources when instantiated for multiplying $n$\by$n$ matrices down to sizes $n = 32$ and $n = 24$, respectively.
The proposed systolic array architectures increase conventional multiplications/multiplier/clock cycle limits by a factor of $1.14^r$ for $r$ implemented levels of Strassen recursion.
Furthermore, they allow the throughput per clock cycle roof of an accelerator to double for each implemented level of Strassen recursion without increasing the minimum supported matrix sizes that can be efficiently multiplied.

\bibliographystyle{IEEEtran}
\bibliography{IEEEabrv,bibl}

\begin{thebibliography}{10}
\providecommand{\url}[1]{#1}
\csname url@samestyle\endcsname
\providecommand{\newblock}{\relax}
\providecommand{\bibinfo}[2]{#2}
\providecommand{\BIBentrySTDinterwordspacing}{\spaceskip=0pt\relax}
\providecommand{\BIBentryALTinterwordstretchfactor}{4}
\providecommand{\BIBentryALTinterwordspacing}{\spaceskip=\fontdimen2\font plus
\BIBentryALTinterwordstretchfactor\fontdimen3\font minus
  \fontdimen4\font\relax}
\providecommand{\BIBforeignlanguage}[2]{{%
\expandafter\ifx\csname l@#1\endcsname\relax
\typeout{** WARNING: IEEEtran.bst: No hyphenation pattern has been}%
\typeout{** loaded for the language `#1'. Using the pattern for}%
\typeout{** the default language instead.}%
\else
\language=\csname l@#1\endcsname
\fi
#2}}
\providecommand{\BIBdecl}{\relax}
\BIBdecl

\bibitem{fuchs2019accelator}
A.~Fuchs and D.~Wentzlaff, ``The accelerator wall: Limits of chip
  specialization,'' in \emph{Proc. IEEE Int. Symp. High Perform. Comput.
  Archit. (HPCA)}, 2019, pp. 1--14.

\bibitem{liu2021winocnn}
X.~Liu \emph{et~al.}, ``{WinoCNN}: Kernel sharing {Winograd} systolic array for
  efficient convolutional neural network acceleration on {FPGA}s,'' in
  \emph{Proc. IEEE 32nd Int. Conf. Appl.-Specific Syst., Arch. Processors
  (ASAP)}, 2021, pp. 258--265.

\bibitem{jouppi2017datacenter}
N.~P. Jouppi \emph{et~al.}, ``In-datacenter performance analysis of a tensor
  processing unit,'' in \emph{Proc. 44th Annu. Int. Symp. Comput. Archit.
  (ISCA)}, 2017, pp. 1--12.

\bibitem{norrie2021design}
T.~Norrie \emph{et~al.}, ``The design process for {Google's} training chips:
  {TPUv2} and {TPUv3},'' \emph{IEEE Micro}, vol.~41, no.~2, pp. 56--63, 2021.

\bibitem{lavin2016fast}
A.~Lavin and S.~Gray, ``Fast algorithms for convolutional neural networks,'' in
  \emph{Proc. IEEE Conf. Comput. Vision Pattern Recognit. (CVPR)}, 2016, pp.
  4013--4021.

\bibitem{pogue2024fast}
T.~E. Pogue and N.~Nicolici, ``Fast inner-product algorithms and architectures
  for deep neural network accelerators,'' \emph{IEEE Trans. Comput.}, vol.~73,
  no.~2, pp. 495--509, 2024.

\bibitem{strassen1969gaussian}
V.~Strassen, ``Gaussian elimination is not optimal,'' \emph{Numer. Math.},
  vol.~13, no.~4, pp. 354--356, 1969.

\bibitem{sun2023accelerate}
J.~Sun \emph{et~al.}, ``Accelerate dense matrix multiplication on
  heterogeneous-{GPU}s,'' in \emph{2023 IEEE 29th Int. Conf. Parallel and
  Distrib. Syst. (ICPADS)}, 2023, pp. 2726--2729.

\bibitem{misra2022stark}
C.~Misra \emph{et~al.}, ``Stark: Fast and scalable {S}trassen’s matrix
  multiplication using apache spark,'' \emph{IEEE Trans. Big Data}, vol.~8,
  no.~3, pp. 699--710, 2022.

\bibitem{krishnan2021multi}
A.~G. Krishnan and D.~Goswami, ``Multi-stage memory efficient {S}trassen's
  matrix multiplication on {GPU},'' in \emph{2021 IEEE 28th Int. Conf. on High
  Perform. Comput., Data, and Anal. (HiPC)}, 2021, pp. 212--221.

\bibitem{khaled2020applying}
A.~Khaled \emph{et~al.}, ``Applying fast matrix multiplication to neural
  networks,'' in \emph{Proc. 35th Annu. ACM Symp. on Appl. Comput.}, 2020, pp.
  1034--1037.

\bibitem{huang2020strassen}
J.~Huang \emph{et~al.}, ``Strassen’s algorithm reloaded on {GPU}s,''
  \emph{ACM Trans. on Math. Softw. (TOMS)}, vol.~46, no.~1, pp. 1--22, 2020.

\bibitem{lai2013accelerating}
P.-W. Lai \emph{et~al.}, ``Accelerating {S}trassen-{W}inograd's matrix
  multiplication algorithm on {GPU}s,'' in \emph{IEEE 20th Annu. Int. Conf. on
  High Perf. Comput}, 2013, pp. 139--148.

\bibitem{lipshitz2012communication}
B.~Lipshitz \emph{et~al.}, ``Communication-avoiding parallel {S}trassen:
  Implementation and performance,'' in \emph{Proc. Int. Conf. High Perform.
  Comput. Netw. Storage Anal.}, 2012, pp. 1--11.

\bibitem{ballard2012communication}
G.~Ballard \emph{et~al.}, ``Communication-optimal parallel algorithm for
  {S}trassen's matrix multiplication,'' in \emph{Proc. 24th Annu. ACM Symp.
  Parallelism Algorithms Archit.}, 2012, pp. 193--204.

\bibitem{winograd1971multiplication}
S.~Winograd, ``On multiplication of 2$\times$ 2 matrices,'' \emph{Linear
  Algebra Appl.}, vol.~4, no.~4, pp. 381--388, 1971.

\bibitem{jouppi2023tpu}
N.~Jouppi \emph{et~al.}, ``{TPU} v4: An optically reconfigurable supercomputer
  for machine learning with hardware support for embeddings,'' in \emph{Proc.
  50th Annu. Int. Symp. Comput. Archit. (ISCA)}, 2023, pp. 1--14.

\bibitem{zhang2019caffeine}
C.~Zhang \emph{et~al.}, ``Caffeine: Toward uniformed representation and
  acceleration for deep convolutional neural networks,'' \emph{IEEE IEEE Trans.
  Comput.-Aided Design Integr. Circuits Syst.}, vol.~38, no.~11, pp.
  2072--2085, 2019.

\bibitem{kung2019maestro}
H.~T. Kung \emph{et~al.}, ``Maestro: A memory-on-logic architecture for
  coordinated parallel use of many systolic arrays,'' in \emph{Proc. IEEE 32nd
  Int. Conf. Appl.-Specific Syst., Arch. Processors (ASAP)}, vol. 2160-052X,
  2019, pp. 42--50.

\bibitem{yuzuguler2023Scale}
A.~C. Y\"{u}z\"{u}g\"{u}ler \emph{et~al.}, ``Scale-out systolic arrays,''
  \emph{ACM Trans. Archit. Code Optim.}, vol.~20, no.~2, mar 2023.

\bibitem{elfimova2001fast}
L.~Elfimova and Y.~V. Kapitonova, ``A fast algorithm for matrix multiplication
  and its efficient realization on systolic arrays,'' \emph{Cybern. Syst.
  Anal.}, vol.~37, no.~1, pp. 109--121, Jan. 2001.

\bibitem{leon2023accel}
L.~G. León-Vega \emph{et~al.}, ``Acceleration of fully connected layers on
  {FPGA} using the {S}trassen matrix multiplication,'' in \emph{Proc. IEEE 5th
  Int. Conf. BioInspired Processing (BIP)}, 2023, pp. 1--6.

\bibitem{ffip-source}
\BIBentryALTinterwordspacing
T.~E. Pogue and N.~Nicolici, ``{FFIP} accelerator implementation,'' 2024.
  [Online]. Available: \url{https://github.com/trevorpogue/algebraic-nnhw}
\BIBentrySTDinterwordspacing

\bibitem{Lakshmi2022}
V.~Lakshmi \emph{et~al.}, ``A novel in-memory wallace tree multiplier
  architecture using majority logic,'' \emph{IEEE Trans. Circuits Syst. I},
  vol.~69, no.~3, pp. 1148--1158, 2022.

\bibitem{guo2019dl}
K.~Guo \emph{et~al.}, ``[{DL}] a survey of {FPGA}-based neural network
  inference accelerators,'' \emph{ACM Trans. Reconfigurable Technol. Syst.},
  vol.~12, no.~1, pp. 1--26, 2019.

\bibitem{Pekmestzi1999}
K.~Pekmestzi, ``Multiplexer-based array multipliers,'' \emph{IEEE Trans.
  Comput.}, vol.~48, no.~1, pp. 15--23, 1999.

\bibitem{intel-dsp}
\BIBentryALTinterwordspacing
``Intel {Arria} 10 native fixed point {DSP IP} core user guide,'' 2024.
  [Online]. Available:
  \url{https://www.intel.com/content/www/us/en/docs/programmable/683583/current/intel-arria-native-fixed-point-dsp-ip.html}
\BIBentrySTDinterwordspacing

\bibitem{kaiming2016deep}
K.~He \emph{et~al.}, ``Deep residual learning for image recognition,'' in
  \emph{Proc. IEEE Conf. Comput. Vision Pattern Recognit. (CVPR)}, 2016, pp.
  770--778.

\bibitem{sx-dev-kit}
``Intel {Arria 10 SoC} development kit,'' 2024.

\bibitem{liu2022toward}
S.~Liu \emph{et~al.}, ``Toward full-stack acceleration of deep convolutional
  neural networks on {FPGA}s,'' \emph{IEEE Trans. Neural Netw. Learn. Syst.},
  vol.~33, no.~8, pp. 3974--3987, 2022.

\bibitem{fan2022fpga}
H.~Fan \emph{et~al.}, ``{FPGA}-based acceleration for bayesian convolutional
  neural networks,'' \emph{IEEE Trans. Comput.-Aided Design Integr. Circuits
  Syst.}, vol.~41, no.~12, pp. 5343--5356, 2022.

\bibitem{huang2022fpga}
W.~Huang \emph{et~al.}, ``{{FPGA-based}} high-throughput {{CNN}} hardware
  accelerator with high computing resource utilization ratio,'' \emph{IEEE
  Trans. Neural Netw. Learn. Syst.}, vol.~33, no.~8, pp. 4069--4083, 2022.

\bibitem{kim2023agamotto}
D.~Kim \emph{et~al.}, ``Agamotto: A performance optimization framework for
  {{CNN}} accelerator with row stationary dataflow,'' \emph{IEEE Trans.
  Circuits Syst. I}, vol.~70, no.~6, pp. 2487--2496, 2023.

\bibitem{amd-dsp}
\BIBentryALTinterwordspacing
``{UltraScale} architecture {DSP} slice,'' 2024. [Online]. Available:
  \url{https://docs.amd.com/v/u/en-US/ug579-ultrascale-dsp}
\BIBentrySTDinterwordspacing

\bibitem{an2022opencl}
J.~An \emph{et~al.}, ``An {OpenCL}-based {FPGA} accelerator for {Faster
  R-CNN},'' \emph{Entropy}, vol.~24, no.~10, p. 1346, 2022.

\bibitem{dai2024dcp}
K.~Dai \emph{et~al.}, ``{DCP-CNN}: Efficient acceleration of {CNNs} with
  dynamic computing parallelism on {FPGA},'' \emph{IEEE Trans. Comput.-Aided
  Design Integr. Circuits Syst.}, 2024.

\end{thebibliography}

\begin{IEEEbiography}[{\includegraphics[width=1in,height=1.25in,clip,keepaspectratio]{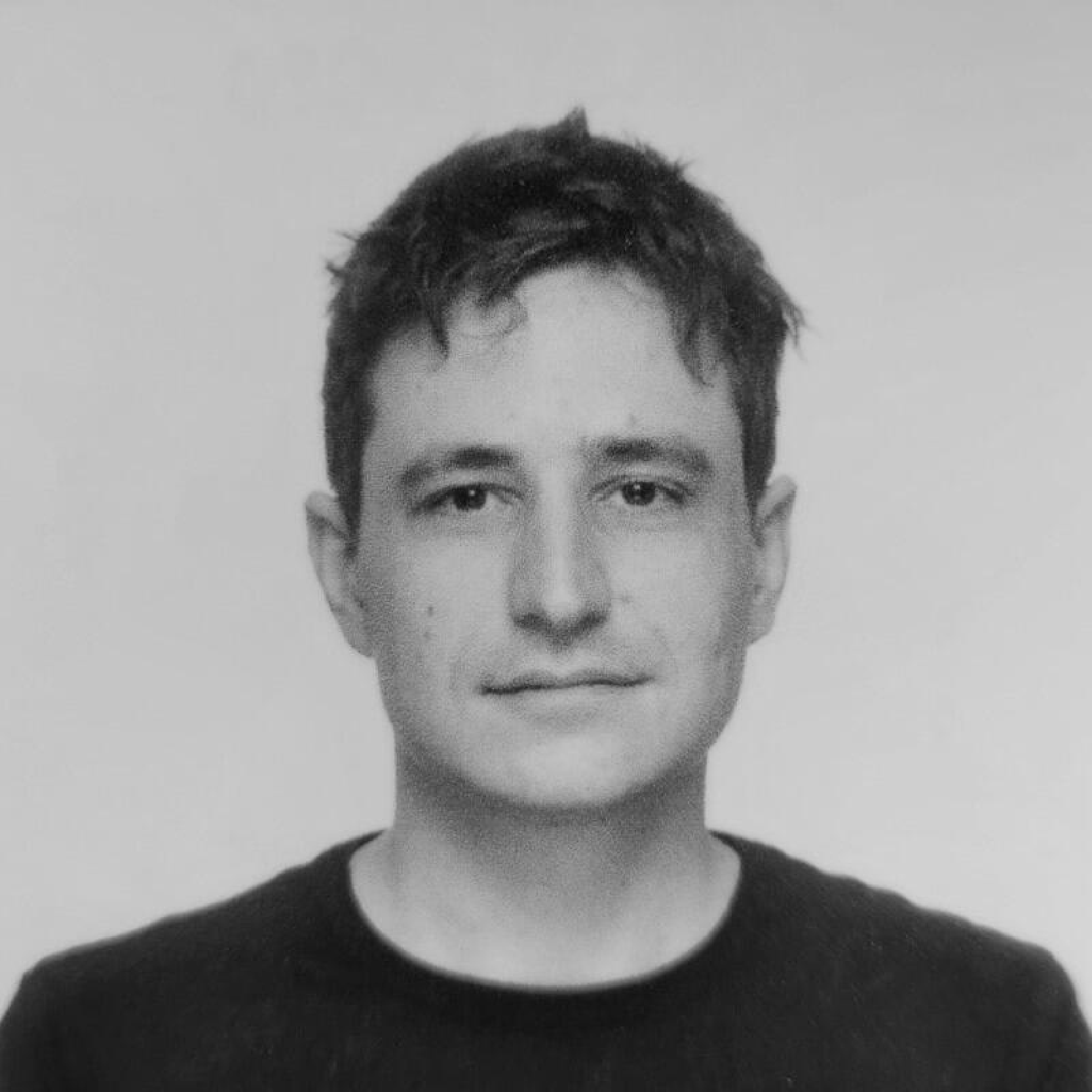}}]{Trevor E. Pogue}
  Trevor E. Pogue received the B.Eng. degree in Electrical Engineering and the M.A.Sc. degree in Electrical and Computer Engineering from McMaster University, Hamilton, Canada, in 2016 and 2019, respectively. He worked as an intern at Synopsys and AMD in 2018 and 2022-2023, respectively. He is currently a Ph.D. Candidate in the Department of Electrical and Computer Engineering at McMaster University, Hamilton, Canada. His research interests are in the area of hardware acceleration.
\end{IEEEbiography}

\begin{IEEEbiography}[{\includegraphics[width=1in,height=1.25in,clip,keepaspectratio]{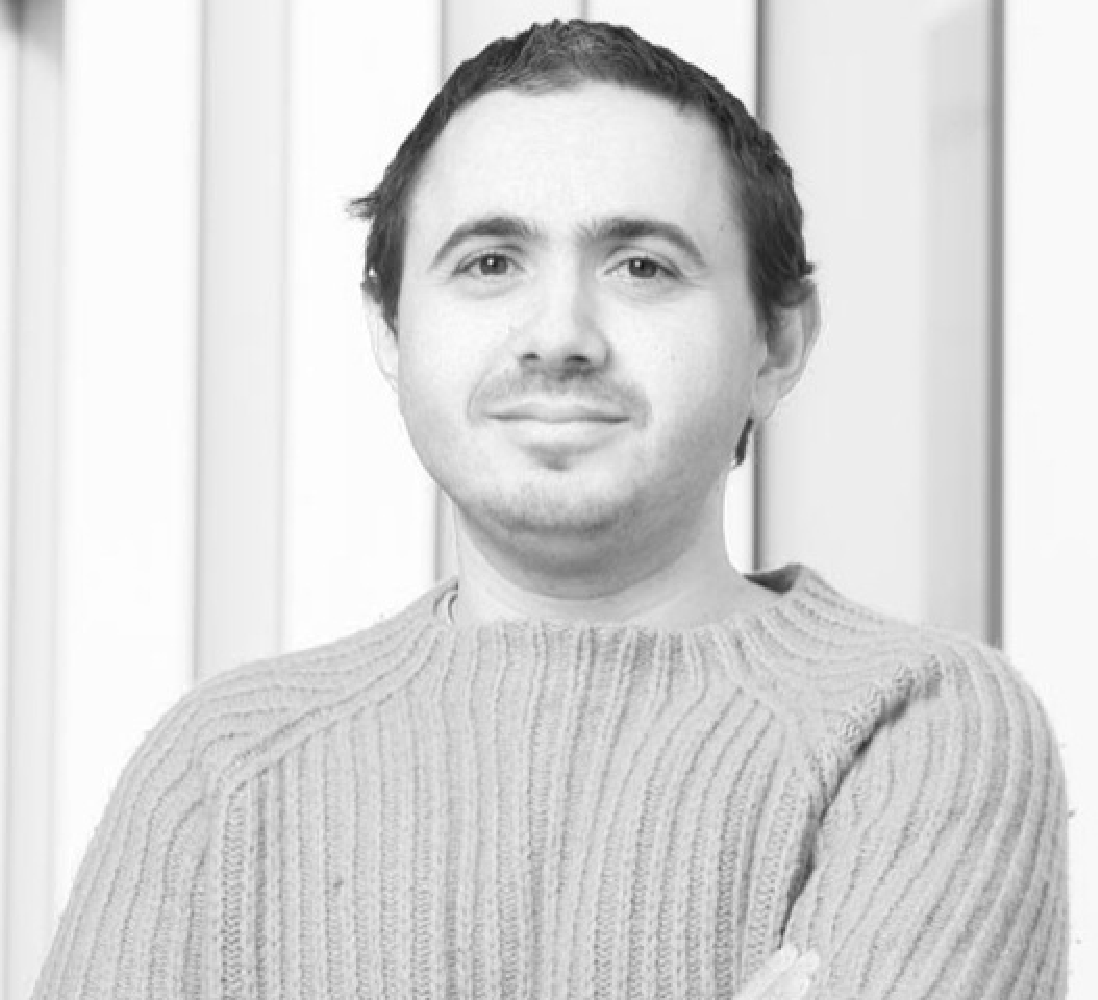}}]{Nicola
    Nicolici}(S’99-M’00-SM'11)
  Nicola Nicolici (S99-M00-SM’11) received the Dipl.Ing. degree in Computer
  Engineering from the “Politehnica” University of Timisoara, Romania, in 1997
  and the Ph.D. degree in Electronics and Computer Science from the University
  of Southampton, U.K., in 2000. He is currently a Professor with the Department
  of Electrical and Computer Engineering, McMaster University, Hamilton, Canada.
  His research interests are in the area of computer-aided design and test. He
  has authored a number of papers in this area. Dr. Nicolici was the recipient
  of the IEEE TTTC Beausang Award for the Best Student Paper at the
  International Test Conference in 2000 and the Best Paper Award at the IEEE/ACM
  Design Automation and Test in Europe Conference in 2004.
\end{IEEEbiography}

\end{document}